\documentclass[prd,twocolumn,nofootinbib,aps,tightenlines,preprintnumbers,notitlepage,
superscriptaddress,floatfix]{revtex4-2}
\usepackage[dvipsnames]{xcolor}
\usepackage{color}
\usepackage{tikz}
\usepackage[caption=false]{subfig}
\usepackage{hyperref}
\usepackage{multirow}
\usepackage{siunitx}
\usepackage[export]{adjustbox}
\usepackage{graphicx}
\usepackage{dcolumn}
\usepackage{bm}
\usepackage[normalem]{ulem}
\usepackage{epstopdf}
\usepackage{tabularx}
\usepackage{suffix}
\usepackage{mathtools}
\usepackage{graphicx}
\usepackage{dcolumn}
\usepackage{bm}
\usepackage[normalem]{ulem}
\usepackage{xcolor}
\usepackage{epstopdf}

\definecolor{TurkishBlue}{HTML}{144893}
\hypersetup{colorlinks=true,citecolor=TurkishBlue,linkcolor=TurkishBlue,urlcolor=TurkishBlue}

\graphicspath{ {figures/} }

\definecolor{mypink}{HTML}{C34D85}

\def\bk{\boldsymbol{k}}

\def\br{\boldsymbol{r}}
\newcommand{\bn}{\hat{ \mathbf{n}}}
\newcommand{\hatv}{\hat{v}}
\newcommand{\hateta}{\hat{\eta}}
\def\bv{\boldsymbol{v}}
\newcommand{\bl}{{\boldsymbol{\ell}}}
\newcommand{\dd}{{\rm d}}
\newcommand{\td}[1]{{\tilde{#1}}}

\def\VEV#1{{\left\langle #1 \right \rangle}}
\def\lr#1#2#3{{\left#1 #2 \right#3}}

\newcommand{\be}{\begin{eqnarray}}

\newcommand{\ee}{\end{eqnarray}}
\newcommand{\apjl}{Astrophys. J. Lett.}
\newcommand{\aap}{Astron. Astrophys.}

\newcommand{\aj}{Astron. J.}

\newcommand{\mnras}{Mon. Not. R. Astron. Soc.}

\newcommand{\Plin}{P^{\rm lin}_{mm}}
\newcommand{\Pion}{P^{\rm ion}_{ee}}
\newcommand{\bxH}{\bar{x}_{\rm H}}
\newcommand{\xHe}{x_{\rm He}}
\newcommand{\xH}{x_{\rm H}}
\newcommand{\bxHe}{\bar{x}_{\rm He}}
\newcommand{\xtot}{x_{\rm tot}}
\newcommand{\bxtot}{\bar{x}_{\rm tot}}

\newcommand{\neion}{n_e^{\rm ion}}
\newcommand{\netot}{n_e^{\rm tot}}
\newcommand{\bneion}{\bar{n}_e^{\rm ion}}
\newcommand{\bnetot}{\bar{n}_e^{\rm tot}}

\newcommand{\Nellbins}{N_{\ell\textrm{-bins}}}
\newcommand{\Nzbins}{N_{z\textrm{-bins}}}

\newcommand{\jhu}{William H. Miller III Department of Physics and Astronomy, Johns Hopkins University, 3400 N Charles St, Baltimore, MD 21218, USA}

\newcommand{\lbnl}{Lawrence Berkeley National Laboratory, One Cyclotron Road, Berkeley, CA 94720, USA}

\newcommand{\berkeley}{Berkeley Center for Cosmological Physics, Department of
Physics, University of California, Berkeley, CA 94720, USA}

\newcommand{\perimeter}{Perimeter Institute for Theoretical Physics, 31 Caroline St N, Waterloo, ON N2L 2Y5, Canada}

\begin{document}

\title{Patchy Helium and Hydrogen Reionization from the \\ Kinetic Sunyaev-Zel’dovich Effect and Galaxies}
\author{Neha~Anil~Kumar}
\email{nanilku1@jhu.edu}
\affiliation{\jhu}

\author{Mesut~\c{C}al{\i}\c{s}kan}
\email{caliskan@jhu.edu}
\affiliation{\jhu}

\author{Selim~C.~Hotinli}
\email{shotinli@perimeterinstitute.ca}
\affiliation{\perimeter}

\author{Marc~Kamionkowski}
\affiliation{\jhu}

\author{Simone Ferraro}
\affiliation{\lbnl}
\affiliation{\berkeley}

\author{Kendrick Smith}
\affiliation{\perimeter}

\date{\today}


\begin{abstract}
Upcoming cosmic microwave background (CMB) experiments will measure temperature fluctuations on small angular scales with unprecedented precision, enabling improved measurements of the kinetic Sunyaev-Zel'dovich (kSZ) effect. This secondary anisotropy has emerged as a valuable probe of the distribution of ionized electrons in the post-recombination Universe. Although the sensitivity of the kSZ effect has recently been utilized to study the high-redshift epoch of hydrogen (H) reionization, its redshift-integrated nature---combined with anticipated improvements in measurement precision---suggests that accounting for the later epoch of helium (He) reionization will become increasingly important in the near future. Joint characterization of the epochs will allow for a more coherent understanding of early-star and -quasar formation, as these sources drive the ionization of H and He in the intergalactic medium. In this paper, we extend the kSZ higher-order statistic introduced by Smith \& Ferraro (2017) to forecast the ability of upcoming CMB surveys to probe the morphology of both H and He reionization. Moreover, given that upcoming large-scale structure surveys will trace density fluctuations at redshifts overlapping with the epoch of He reionization, we propose a novel cross-correlation between the kSZ higher-order statistic and galaxy survey measurements. Using a joint information-matrix analysis of H and He reionization,
we show that next-generation CMB and galaxy surveys will have sufficient statistical power to characterize the patchy morphology of H reionization and set constraints on the redshift evolution of its He counterpart.
\end{abstract}

\maketitle


\section{Introduction}
Secondary effects embedded in the observed cosmic microwave background (CMB) signal, arising from interactions between the CMB photons and free electrons along the line of sight, are known to carry a trove of information regarding the evolution of the post-recombination Universe.
Therefore, leveraging these effects offers a powerful means of characterizing two of the most significant transitions in the morphology of the intergalactic medium (IGM): the reionization of hydrogen (H) and helium (He).
The former epoch of H reionization~\citep[e.g.,][for reviews]{Barkana:2000fd, Pritchard:2011xb}, has been studied in considerable detail thanks to a wealth of observational evidence~\citep[e.g.,][]{Planck:2015fie,Robertson:2013bq,Fan:2006dp,Fan:2005es} and advanced theoretical modeling~\citep[e.g.,][]{Murray:2020trn}. 
In contrast, however, its He counterpart has received less attention largely due to is low relative abundance (8\% that of H) leading to a paucity of direct observables and more subtle observational signatures.
Despite this difference, theoretical models of and forecasts on the morphology of this later epoch~\citep[e.g.,][]{Furlanetto:2007gn, McQuinn2009, Kapahtia:2024rgw, Hotinli:2022jna, Hotinli:2022jnt, Caliskan:2023yov, LaPlante:2015rea, LaPlante:2016bzu, LaPlante:2017xzz,LaPlante2017,LaPlante2018}, and a few observational signatures~\citep[e.g.,][]{Boera2016,Bolton2017,Hiss:2017qyw,Walther:2018pnn,Gaikwad:2020art,Gaikwad:2020eip} have demonstrated that investigating both H and He reionization in tandem holds the promise of a more coherent picture of cosmic reionization, offering deeper insights into the sources of ionizing photons. 

Given this motivation, in this paper, we use the statistical technique developed in Refs.~\cite{Smith:2016lnt, Ferraro:2018izc}---originally used to constrain H reionization---to propose a novel cross-correlation with galaxy survey data, aiming to jointly characterize the patchy morphology of both the epochs. The original statistic, introduced in Ref.~\cite{Smith:2016lnt}, uses the kinetic Sunyaev-Zel’dovich (kSZ) effect---a secondary CMB anisotropy sourced by the scattering of photons off free electrons in the IGM moving with non-zero bulk velocity relative to the CMB rest frame.
Because the epoch of H reionization involves a gradual, finite and anisotropic change in the free-electron fraction, driven by stellar radiation at $z\sim 8.5$, the kSZ effect was identified as a potential probe of the mean redshift and duration of the epoch. However, recent attempts to use this dataset to probe H reionization have shown that extracting precise constraints remains challenging~\citep[see e.g.,][and references therein]{Jain:2023jpy}.
This challenge is largely attributed to the fact that, at the power-spectrum level, it is difficult to distinguish the H reionization-sourced kSZ signal from that produced by anisotropic distributions of ionized electrons surrounding low-redshift galaxies.  
In order to alleviate this degeneracy, the authors of Ref.~\cite{Smith:2016lnt} proposed a higher-order (trispectrum-level) statistic that was shown to isolate the kSZ signal and effectively separate the late-time and H-reionization kSZ. 
Forecasts using this technique from Ref.~\cite{Ferraro:2018izc} have indicated that upcoming CMB surveys should be able to constrain the optical depth to recombination $\bar{\tau}$, as well as the expected duration of reionization well within $\sim 3\sigma$. Those forecasts were extended to include information from the kSZ power spectrum by Ref.~\cite{Alvarez:2020gvl}.
This method was also recently applied to observed CMB data from the South Pole Telescope~\cite{SPT-3G:2024lko} and the Atacama Cosmology Telescope~\cite{MacCrann:2024ahs}, with the latter resulting in an upper limit on detection of the trispectrum signal and the former quoting an upper bound on the expected duration of the epoch.

Unlike H reionization, the second ionization of He requires photons with energies exceeding $\sim54$ eV, well beyond the photon-energies emitted by early populations of stars.
Consequently, He remained singly ionized until $z\sim3$, when high-energy photons from quasars and active galactic nuclei (AGN) could initiate its complete reionization~\citep[e.g.,][]{Villasenor:2021ksg,Gaikwad:2020eip,Gaikwad:2020art,Walther:2018pnn,Boera:2018vzq,Hiss:2017qyw,Bolton:2013cba}. 
The morphology and timing of this epoch are, therefore, tightly coupled to the abundance and properties of quasars~\cite{2012ApJ...755..169M, 2013ApJ...773...14R, 2013ApJ...768..105M, McGreer:2017myu, 2022ApJ...928..172P}, the behavior of AGN~\citep{Shen:2014rka, Hopkins:2006vv, 2017ApJ...847...81S}, and the formation and evolution of supermassive black holes~\citep{Inayoshi:2019fun}. 
Thus, joint modeling of H and He reionization offers a unique opportunity to disentangle the respective roles of stellar populations and quasars as ionizing sources, with the interplay between the two epochs offering insights into the thermal evolution of the IGM and timing of structure formation~\citep[e.g., see][]{Villasenor:2021ksg}.

Given the similarity in the physical processes governing both the epoch of H and He reionization, as well as their separation in redshift space, a natural step towards joint detection is incorporating the effects of He reionization into the model of the trispectrum signal from Ref.~\cite{Smith:2016lnt}. 
However, joint modeling and forecasts from Ref.~\cite{Caliskan:2023yov} demonstrate that detecting a characteristic signal from He reionization using a redshift-integrated probe such as the CMB alone is likely unfeasible, due to the low relative abundance of He.
Reference~\cite{Caliskan:2023yov} further shows that incorporating low-redshift cross-correlations can substantially enhance the sensitivity to He reionization.
This intuition ultimately motivates our proposal to cross-correlate the kSZ data set with galaxy survey measurements.

In this paper, we first extend the model of the proposed trispectrum signal from Ref.~\cite{Smith:2016lnt} to incorporate not only the effects of `patchy' H reionization---including parameters that determine the size distribution of ionized regions---but also a dependence on the He signal.
This involves propagating an existing analytical model of reionization (used in Refs.~\cite{Dvorkin:2008tf, Caliskan:2023yov}) to its impact on the distribution of ionized electrons.
In our derivation, we now include a direct dependence on the assumed small-scale total electron distribution, allowing for forecasts that may further instruct on how halo-occupation-distribution (HOD) model assumptions affect reionization-parameter errors. 
Based on this model, we then compute the expected kSZ-based trispectrum signal and noise.
Moreover, anticipating a lack of sensitivity from the $z$-integrated trispectrum to He reionization, we also present a novel cross-correlation between the kSZ-weighted temperature squared maps used in Ref.~\cite{Smith:2016lnt} and galaxy survey measurements.
Since galaxy surveys trace the \textit{three-dimensional} distributions of ionized electrons sourcing the late-time kSZ effect, it can be used as a tracer of the cosmological radial velocity field (squared) that induces large-scale variations in the kSZ local power. 
Therefore, the combination of these two data sets allows for the construction of a suite of redshift-binned cross-correlation signals at low redshifts, overlapping with the epoch of He reionization and amplifying its effects relative to its H counterpart.

Given the set of signals elucidated above, we forecast the expected detection signal-to-noise ratio (SNR) for both the total auto-correlation and the proposed cross-correlation signals. 
We also forecast the future measurability of the morphology of each epoch by presenting results from our information-matrix analysis.
In our analyses, we \textit{jointly} forecast parameter errors on characteristics of both H and He reionization, ensuring that any degeneracy between the two epochs is accounted for.
These forecasts are first computed using only the auto-correlation trispectrum to demonstrate its weaker sensitivity to He reionization.
This is followed by parameter-error forecasts with the cross-correlation signal accounted for, clearly displaying the improvement driven by the inclusion of redshift-binned signals at $z \lesssim 5$.
All the presented forecasts are computed for two separate baselines: the first corresponding to a CMB-S4-like survey cross-correlated with galaxy-survey data from LSST, and the second similarly constructed using experiment specifications matching CMB-HD, cross-correlated with the upcoming MegaMapper\footnote{Since the original ``MegaMapper'' concept has evolved into the broader Stage-5 spectroscopic survey (Spec-S5) framework, we take MegaMapper to represent a plausible Spec-S5-like survey~\cite{Spec-S5:2025uom}.} survey. 

This paper is organized as follows. In Sec.~\ref{sec: Reionization Model for the Ionized Electron Power Spectrum}, we derive our model for the ionized electron power spectrum, including a description of the model characterizing patchy reionization. 
In Sec.~\ref{sec: kSZ Trispectrum and Galaxy Cross-Correlation}, we briefly summarize the derivation of the kSZ-based trispectrum proposed by Ref.~\cite{Smith:2016lnt} and compute this $z$-integrated auto-correlation signal assuming our chosen model of reionization.
We follow-up this discussion with the proposal for cross-correlation with galaxy-survey measurements. 
Here, we derive expressions characterizing the $z$-binned cross-correlation signal and noise. 
The experiment specifications characterizing each baseline and the associated forecasts are presented in Sec.~\ref{sec:Forecasts}.
We start by describing the survey specifications and assumed sources of noise and foregrounds for each upcoming experiment considered in our forecasts. 
We then present the forecasted detection-SNR for the auto- and cross-correlation signals from each baseline. 
Finally, to display the power of the cross-correlation probe, we present fractional-error forecasts on the parameters characterizing both H and He reionization, computed jointly using our information-matrix analysis.
We conclude with a discussion in Sec.~\ref{sec: Conclusions}. 
Throughout the paper, we assume a $\Lambda$CDM cosmology with cosmological parameter values consistent with Planck 2018~\cite{Aghanim:2018eyx}. 
Unless specified otherwise, we work under the convention $c = 1$.

\section{Reionization Model for Ionized Electron Power Spectrum}
\label{sec: Reionization Model for the Ionized Electron Power Spectrum}
In this section, we present our model for the reionization of both He and H, and outline how it translates into a prescription for the redshift evolution of the ionized electron power spectrum.
Our analytical model of the ionized electron power spectrum involves an extension of the standard HOD model, accounting for both 2-halo (2$h$) and 1-halo (1$h$) calculations~\cite{2011ApJ...738...45L}. 

We begin by setting up the base model for the distribution of ionized electrons, which incorporates anisotropies in the \textit{total} electron distribution (regardless of ionization state) as well as fluctuations in the ionization field. 
Next, we establish our HOD-based assumptions regarding the distribution of total electrons in the IGM. 
We then describe our model for reionization and extend this HOD-based framework to construct the relevant anisotropic ionization field in Fourier space, drawing from the derivations presented in Ref.~\cite{Mortonson:2006re}. 
All expressions are initially presented in a model-agnostic format, highlighting their general dependence on the assumed electron distribution profile, the expected distribution of ionized regions, and the redshift evolution of the average ionization fraction. 
We then conclude with a final expression for the ionized electron power spectrum summarized in Eqs.~\eqref{eq: ion_elec_PS_as_ft_2pt_terms}--\eqref{eq: ne_tot_x_X_fourier_two_point}.

\subsection{Reionization \& the Distribution of Electrons}
\label{subsec: Reionization & the Distribution of Electrons}
The comoving number-density of \textit{ionized} electrons can be expressed as the \textit{total} (ionized + non-ionized) electron number density $\netot(\br,z)$ modulated by the ionization fraction field $\xtot(\br, z)$ as follows:
\begin{eqnarray}
    \neion(\br) &= &\xtot(\br)\netot(\br)\,, \nonumber\\
    &= &\left[(1 - \epsilon)\xH(\br) +  \epsilon\xHe(\br)\right] \netot(\br)\,,
    \label{eq: ionized_elec_base_model}
\end{eqnarray}
where the $z$-dependence of all the fields and distributions has been suppressed for ease of notation. 

Moreover, we assume that spatial fluctuations in the $\neion(\br, z)$ distribution are sourced by a combination of H-ionized electrons as well as those from the \textit{second} ionization of He. 
As a result, in the second line of Eq.~\eqref{eq: ionized_elec_base_model}, we expand $\xtot(\br)$ into its two contributors: one accounting for ionization of H in the IGM $\xH(\br)$ and the other accounting for the second ionization of He $\xHe(\br)$. 
Each of these independent ionization fields is defined as follows:
\begin{eqnarray}
    x_X(\br) = \begin{cases} 0 & \text{if $X$ is neutral at location $\br$}\,, \\ 
    1 & \text{if $X$ is ionized at location $\br$}\,,\end{cases}
\end{eqnarray}
where $X \in \{{\rm H},\, {\rm He}\}$. 
In this model, if $X = {\rm He}$, the ionization field can only `act on' singly-ionized He atoms. 
Therefore, given the primordial He abundance $Y_p$, the field $\xHe(\br)$ can only account for (up-to) $\epsilon \equiv (Y_p/4)/(1 - Y_p/2)$ of the total number of ionized electrons. 
Similarly, if $X = {\rm H}$, the field can only account for the ionization of (up-to) $(1-\epsilon)$ of the total number of non-ionized electrons. 
This separation of the total non-ionized electron population into two separate `species' then allows for an ionization fraction field $\xtot(\br)$ that is unit normalized. 

Given the expression for the ionized electron distribution in Eq.~\eqref{eq: ionized_elec_base_model}, the two point correlation function of the distribution (in real-space) is given by:
\begin{align}
    \VEV{\neion\neion}_r &= \bxtot^2 \VEV{\netot\netot}_r 
    + \left(\bnetot\right)^2 \VEV{\xtot\xtot}_r \nonumber\\
    &\quad+ 2\bxtot\bnetot \VEV{\netot\xtot} \nonumber\\
    &\quad+ \VEV{(\delta\xtot\cdot\delta\netot)(\delta\xtot\delta\netot)}_r\,,
    \label{eq: ion_elec_two_point_terms_real_space}
\end{align}
where we have used the simplifying notation $\VEV{\delta F(\br_1)\delta G(\br_2)} \equiv \VEV{FG}_r$ with $|\br_1 - \br_2| = r$. Furthermore, we have used Wick's theorem to neglect any terms of the form $\VEV{X(\delta\xtot\delta\netot)}_r$ for $X\in\{\xtot, \netot\}$.
Then, the ionized electron power spectrum can be obtained from the above expression as follows:
\begin{eqnarray}
    \Pion(k) &\equiv &\frac{1}{\left[\bneion(z)\right]^2}\VEV{\neion\neion}_k\,, \nonumber\\ &= 
    &\frac{1}{\left[\bneion(z)\right]^2}\int \dd^3\br \VEV{\neion\neion}_re^{i\bk\cdot\br}\,,
    \label{eq: ion_elec_PS_as_ft}
\end{eqnarray}
where $\bneion(z) = \bxtot(z)\bnetot$, and $\bxtot(z)$ and $\bnetot$ are the average ionization fraction and average total-electron (comoving) number density (respectively) at redshift $z$.  

The simplified expression in Eq.~\eqref{eq: ionized_elec_base_model} allows for us to independently model both H and He reionization while also accounting for a possible period of overlap between the two epochs. 
Furthermore, given the above set-up, the expression describing the redshift-evolution of the ionized electron power-spectrum can be calculated by characterizing $\netot(\br, z)$ and a similar, HOD-like prescription for $\xH(\br,z)$ and $\xHe(\br,z)$ (following Ref.~\cite{Mortonson:2006re}).

\subsection{HOD Model for Total Electron Distribution}
\label{subsec: HOD Model for Total Electron Distribution}
Following the HOD prescription, the Fourier-space distribution of \textit{all} electrons (ionized and non-ionized) at redshift $z$ and wave-vector $\bk$ is given by:
\begin{eqnarray}
    \delta\netot(\bk, z) = \int \dd M\ n(M, z)N_e(M,z)u_e(\bk | M, z), 
    \label{eq: tot_elec_dist_fourier}
\end{eqnarray}
where $M$ labels the halo mass, $n(M, z)$ is the halo-mass function (i.e., the differential number-density of halos per unit mass at redshift $z$), $N_e(M, z)$ is the average number of electrons in a halo, and $u_e(\bk | M, z)$ is the normalized Fourier transform of the electron profile. 
Assuming that all electrons are found in halos, and the baryon-to-dark matter ratio within each halo is always the same as the cosmic average, 
\begin{eqnarray}
     N_e(M, z) = \frac{\Omega_b}{\Omega_m}\frac{M}{m_p}\left(1 - Y_p/2\right),
\end{eqnarray}
where $\Omega_m$ and $\Omega_b$ are the matter- and baryon-density parameters today (respectively), and $m_p$ labels the mass of a proton.
Then, the $\bnetot$ can be expressed as
\begin{eqnarray}
    \bnetot =  \frac{\Omega_b}{\Omega_m}\frac{\rho_m}{m_p} \left(1 - Y_p/2\right), 
    \label{eq: average_total_elec_num_density}
\end{eqnarray}
where $\rho_m$ is the matter density today. 
Note that we neglect any deficit in large-scale power caused by the collapse of gas into stars within halos.
In other words, the above formalism is set up such that the total electron power spectrum, on large scales, approximates to the linear matter power spectrum $P_{mm}^{\rm lin}(k,z)$. 

To capture the dependence of our forecasts on the small-scale distribution of electrons ($1h$-scale), we explore three different spherically-symmetric models for $u_e(k | M, z)$. 
In one case, we assume that the electrons are distributed according to an NFW function at sub-halo scales. 
This model leads to a relatively high power on small-scales. 
The second model we consider is the simulation-based AGN model for electron distribution presented in Ref.~\cite{Battaglia:2016xbi}. 
This feedback model predicts a lower power than the NFW scenario. 
Finally, we also explore a model that we label `$W_e(k)\times$NFW', which is an approximation of the small-scale electron distribution obtained by multiplying the non-linear matter power spectrum (also obtained using an NFW model on sub-halo scales) by a window function $W_{e}(k,z)$. 
Reference~\cite{Smith:2016lnt} explains that this fitting-function results in a late-time kSZ power spectrum that approximately agrees with the `cooling+star-formation' model from Ref.~\cite{Shaw:2012abc}. 
This model predicts the lowest power at $k \gtrsim 1\ {\rm Mpc}^{-1}$.
All three models deviate from $P_{mm}^{\rm lin}(k,z)$ on small scales.

The functional forms of the halo-mass function~\cite{Tinker:2008ff}, and each type of electron profile are summarized in Appendix~\ref{appendix: Total Electron Distribution Profile Specifications}. 
Figure~\ref{fig:total_elec_PS_no_re} displays the \textit{total} electron power spectrum ($\propto \VEV{\netot\netot}_k$) for each of the above mentioned models of $u_e(\bk | M, z)$ at $z = 1.0$ calculated according to the summary presented in Appendix~\ref{appendix: Total Electron Distribution Profile Specifications}. 

\subsection{HOD Model for Ionization Fraction Fields}
\label{subsec: HOD Model for Ionization Fraction Fields}
To model the spatial fluctuations in the two independent ionization fraction fields $\xH(\br, z)$ and $\xHe(\br,z)$, we use a methodology previously established in Ref.~\cite{Mortonson:2006re}, that has been utilized in other forecasts on both the epochs~\citep[e.g.,][]{Caliskan:2023yov, Dvorkin:2008tf}. Since both epochs are modeled using the same formalism, we present the parametrization below for a `species' of neutral particles labeled $X$, where  $X \in \{{\rm H},\, {\rm He}\}$.\\ 

\subsubsection{Generalized Model for Patchy Ionization Fraction}
In this parametrization, we assume that $X$ becomes ionized in spherical-bubble like regions surrounding ionizing sources, with full ionization within the radius of the bubble and neutral $X$ outside. Furthermore, we assume that these bubbles populate some large cosmological volume $V_0(z)$ as a Poisson process. As was done in Ref.~\cite{Mortonson:2006re}, averaging over the Poisson process allows us to write the ionization fraction at location $\br$ as:
\begin{eqnarray}
    x_X(\br, z) = 1 - \exp\left[-\int \dd R\ n^X_b(\br, R, z) V_b(R)\right]\,,
    \label{eq: ion_frac_distrib_basic}
\end{eqnarray}
where $n^X_b(\br, R, z)$ is the comoving number density of bubbles with radius $R$, and $V_b(R) \equiv 4\pi R^3/3$ is the volume of each bubble.
Next, we assume that, on large scales, the bubbles trace the linear matter over-density field. This allows us to express the comoving bubble distribution as:
\begin{eqnarray}
    n_b^X(\br, R) = \int\dd M\,P_X(M, R)n(M)[1 + b(M)\delta_m^W(\br)]\,,
    \label{eq: bub_distrib_gen_expr}
\end{eqnarray}
where the redshift dependence of all quantities has been suppressed for ease of notation. 
In the above equation, $P_X(M,R)$ is the probability\footnote{Note that $P_X(M,R)$ is normalized such that $1 - \int \dd R\,P_X(M, R)$ is the probability that a halo of mass $M$ \textit{does not} host a bubble. In other words, $P_X(M,R)$ is not normalized to one.} that a halo of mass $M$ hosts a bubble of radius $R$, $b(M, z)$ is the linear halo-bias, and $\delta_m^W(\br, z)$ is the matter over-density field smoothed by a top-hat window function $W_R(\br)$. The functional forms of $b(M, z)$ and $W_R(\br)$ are detailed in App.~\ref{appendix: Total Electron Distribution Profile Specifications}.

To obtain a generalized expression for the average ionization fraction of particle $X$ at redshift $z$, we can plug Eq.~\eqref{eq: bub_distrib_gen_expr} into Eq.~\eqref{eq: ion_frac_distrib_basic} and expand the variational part of the exponential, before taking an average over cosmological volume $V_0(z)$. This gives us:
\begin{eqnarray}
    \bar{x}_X(z) &= &1 - \exp\lr{[}{-\int \dd M\dd R\,P(M,R)n(M)V_b(R)}{]}\,, \nonumber\\ &\equiv &1 - \exp\lr{[}{-\VEV{n_b^XV_b}(z)}{]}.
    \label{eq: bar_x_generalized}
\end{eqnarray}
The above equation demonstrates that the set of functions $\{\bar{x}_e, n_b^X, V_b\}$ are not all independent, i.e., the functional forms attributed to these physical quantities are subject to the above constraint

Furthermore, given the generalized description of bubbles as a tracer of the matter-density field on large scales, we can also write an expression for the spatial fluctuations in the ionization fraction field of particle $X$ as:
\begin{widetext}
    \begin{eqnarray}
        \delta x_X(\bk, z) = [1 - \bar{x}_X(z)]\int \dd M dR\ P_{X}(M, R, z)V_b(R)n(M,z)b(M,z)W_R(k)\delta_m(\bk, z)\,,
    \label{eq: x_X_fourier_distrib}
    \end{eqnarray}
\end{widetext}
where we have written the expression in terms of $\bar{x}_X(z)$ using Eq.~\eqref{eq: bar_x_generalized}.

\subsubsection{Simplified Model Parametrization}
\label{subsubsec: Simplified Model Parametrization}
As will be clear in the following sections, our estimator, that relies on the kSZ effect to probe the epochs of reionization, is more sensitive to the redshift evolution of the fields than the parametrization of the `patchiness'. 
In terms of the generalized model presented above, this translates to the estimator having a weaker dependence on $P_X(M,R,z)$ and a stronger dependence on $\bar{x}_X(z)$. 
As a result, when characterizing the fields $x_X(\br, z)$, we adhere to simplifying assumptions made in previous work (see, for example, Ref.~\cite{Caliskan:2023yov}). 
Below, we explain these simplifying assumptions and their impact on the generalized equation for $\delta x_X(\bk, z)$ [Eq.~\eqref{eq: x_X_fourier_distrib}]. 
Finally, we establish the parametrization adopted for the final forecasts presented in this paper. 

First, we narrow the parameter space characterizing the distribution of bubbles by assuming that $P_X(M,R,z)$ is independent of redshift $z$, and  any halo with virial mass above a fixed threshold $M_{{\rm th}}^X$ hosts a bubble of radius $R$ with probability $P_X(R)$. 
These assumptions result in the following simplified expression for $\delta x_X(\bk, z)$:
\begin{widetext}
    \begin{eqnarray}
        \delta x_X(\bk, z) = \frac{[1 - \bar{x}_X(z)]\,\ln[1 - \bar{x}_X(z)]}{\VEV{V_b}}b^X(z)\int \dd R\ P_{X}(R)V_b(R)W_R(k)\delta_m(\bk, z)\,,
    \label{eq: x_X_fourier_distrib_simpl}
    \end{eqnarray}
\end{widetext}
where we have defined a new parameter called the bubble bias $b^X(z)$ as
\begin{eqnarray}
    b^{X}(z) &\equiv &\frac{\int_{M_{{\rm th}}^X} \dd M \,n(M,z) b(M,z)}{\int_{M_{\rm th}^X} dM n(M,z)}\,, \nonumber\\ &\equiv &\frac{\int_{M_{{\rm th}}^X} \dd M\, n(M,z) b(M,z)}{\bar{n}_b^X(z)}\,,
\end{eqnarray}
and we have expressed the pre-factor in Eq.~\eqref{eq: x_X_fourier_distrib_simpl} in terms of $\bar{x}_X(z)$ by plugging in:
\begin{eqnarray}
    \bar{x}_X(z) &= &1-\exp\lr{[}{-\bar{n}_b^X(z)\int \dd R\, P_X(R)V_b(R)}{]}\,, \nonumber \\ &\equiv &1-e^{-\bar{n}_b^X(z)\VEV{V_b}}\,.
    \label{eq: bar_x_simplified}
\end{eqnarray}
The above expression for $\bar{x}_X(z)$ is derived from Eq.~\eqref{eq: bar_x_generalized}, accounting for the simplifying assumptions on $P_X(M,R,z)$. 
Our final step of simplification in the following forecasts is then assuming that the bubble bias does not evolve with redshift, i.e., $b^X(z) \approx b^{X}$. 
Therefore, given the constraint in Eq.~\eqref{eq: bar_x_simplified}, we only have to model $\bar{x}_X(z)$ and $P_X(R)$ to compute $\delta x_X(\bk, z)$ and its relevant two-point statistics.

Following Refs.~\cite{Dvorkin:2008tf, Caliskan:2023yov}, we model the redshift-evolution of the average ionization fraction as follows:
\begin{eqnarray}
    \bar{x}_{X}(z)=\frac{1}{2}\left[1-\tanh\left(\frac{y(z)-y_{\rm re}^{X}}{\Delta_y^{X}}\right)\right]\,,
    \label{eq:mean_ionization_fraction_functional_form}
\end{eqnarray}
where $y(z)=(1+z)^{3/2}$, and $y_{\rm re}^X$ and $\Delta_y^X$ are free parameters of the reionization model, effecting the time and duration of the reionization process, respectively. Although the functional form we adopt for the redshift evolution of $\bar{x}_X$ remains the same across both the epochs, we use different fiducial values for the parameters $\{y_{\rm re},\, \Delta_y\}$ depending on whether $X = $ H or $X = $ He.

For H reionization, we assume $\{y_{\rm re}^{\rm H},\, \Delta_y^{\rm H}\} = \{33.5 \,,8.0\}$, (roughly) corresponding to a central redshift of $z\sim 9$ and a duration of $6 \lesssim z \lesssim 12$. 
As explained in Refs.~\cite{Dvorkin:2008tf, Caliskan:2023yov}, this choice is motivated by star-formation astrophysics, given that early stars are most commonly assumed to be the drivers of H reionization. 
The assumed values are consistent with the anticipated properties of the Lyman-$\alpha$ forest~\cite[e.g.,][]{Pritchard:2012abc, Barkana:2000fd}. 
In contrast, for He reionization, we assume $\{y_{\rm re}^{\rm He},\, \Delta_y^{\rm He}\} = \{8.0\,,3.5\}$, which correspond to a mean redshift of $z\sim 3$ and a duration spanning $1 \lesssim z \lesssim 5$. 
As with its H counterpart, the chosen fiducial values are motivated by considering the formation and evolution of galaxies and quasars, given that these high-energy astrophysical bodies are expected to be the main drivers of He reionization~\cite[see~e.g.,][]{Furlanetto:2007gn,McQuinn:2012bq,Worseck:2014gva,Furlanetto:2007mg,Worseck:2011qk,Sokasian:2001xh,Compostella:2013zya,Oh:2000sg,Furlanetto:2007gn,Furlanetto:2008qy,Dixon:2009xa,LaPlante:2016bzu,Caleb:2019apf,Linder:2020aru,Meiksin:2011bq,Compostella:2014joa,Eide:2020xyi,UptonSanderbeck:2020zla,Bhattacharya:2020rtf,Villasenor:2021ksg,Meiksin:2010rv,Gotberg:2019uhh,LaPlante:2015rea,Syphers:2011uw,Dixon:2013gea,LaPlante:2017xzz}. 
Other astrophysical measurements that are consistent with these expectations include, once again, measurements of the Lyman-$\alpha$ forest~\cite[e.g.][]{Miralda-Escude:1993abc, Croft:1997abc, Giroux:1997abc, Bolton:2013cba, Hiss:2017qyw,Boera:2018vzq, Walther:2018pnn, Gaikwad:2020art, Gallego:2021whq} and measurements of the average temperature of the IGM~\cite[e.g.,][]{Rorai:2017qft,Becker:2010cu,Fan:2006dp,Schaye:2000abc, Theuns:2002abc}. 
The chosen set of values are also consistent with the most recent CMB measurements of the optical depth to recombination, setting $\bar{\tau}\approx 0.054$~\cite{Planck:2018vyg}\footnote{Note that the values chosen for $\{y_{\rm re}^X\,, \Delta_y^X\}$ are different from the fiducial values chosen in Ref.~\cite{Caliskan:2023yov}. This is done to account for the difference in normalization of the epoch of H reionization. That is, $\xtot(\br, z)$ can take a maximum value of one, with every occurrence of $\xH(\br, z)$ normalized by $(1-\epsilon) < 1.0$. The values are thus revised to ensure that the optical depth to reionization remains $\bar{\tau}\approx 0.054$ under the updated normalization}. 
However, it is important to note that, due to the low relative abundance of He, we have more freedom in the characterization of $\bxHe(z)$ with minimal impact on the average optical depth of reionization. 

As done with the above model for $\bar{x}_X(z)$, we maintain consistency with existing models of H reionization (from Refs.~\cite{Dvorkin:2008tf, Caliskan:2023yov}) in our approach to characterizing $P_X(R)$.
We assume that the bubble-like volumes surrounding ionizing sources of `species' $X$ have a radial distribution given by a log-normal profile:
\begin{eqnarray}
    P_X(R)=\frac{1}{R}\frac{1}{\sqrt{2\pi(\sigma_{\ln R}^X)^2}}e^{-[\ln(R /\bar{R}^X)]^2/\lr{[}{2(\sigma_{\ln{R}}^X)^2}{]}}\,,
    \label{eq:bubble_radius_distribution_functional_form}
\end{eqnarray}
where $\bar{R}^X$ is the characteristic size of the bubbles and $\sigma_{\ln R}^X$ is the width of the distribution. 
Following Ref.~\cite{Dvorkin:2008tf}, we set $\{\bar{R}^{\rm H},\, \sigma_{\ln R}^{\rm H} \} = \{5\,{\rm Mpc},\, \ln 2\}$ for H reionization. 
In contrast, because we expect He reionization to be driven by luminous quasars or AGN that emit harder photons, we assume that the average size of bubbles in this epoch is $\bar{R}^{\rm He} = 15\,{\rm Mpc}$ (consistent with calculations in Refs.~\cite{Furlanetto:2007gn, Caliskan:2023yov}). 
Due to weaker current constraints and a shortage in simulations of He reionization, we assume $\sigma_{\ln R}^{\rm H} = \sigma_{\ln R}^{\rm He} = \ln 2$. 
It is important to note that although this chosen model of reionization is consistent with previous forecasts on both the epochs, the kSZ effect sourced by this model on small-scales is weaker than the prediction from some simulations (see, for example, Ref.~\cite{2013ApJ...776...83B}). Since the signal from He reionization is much weaker than that of H, the discrepancy is mainly driven by our chosen model for $\xH(\br,z)$. 
For further details on how this affects the proposed trispectrum statistic, see Sec.~\ref{subsec: The kSZ Trispectrum}.

\begin{figure}
    \centering
    \includegraphics[width=\linewidth]{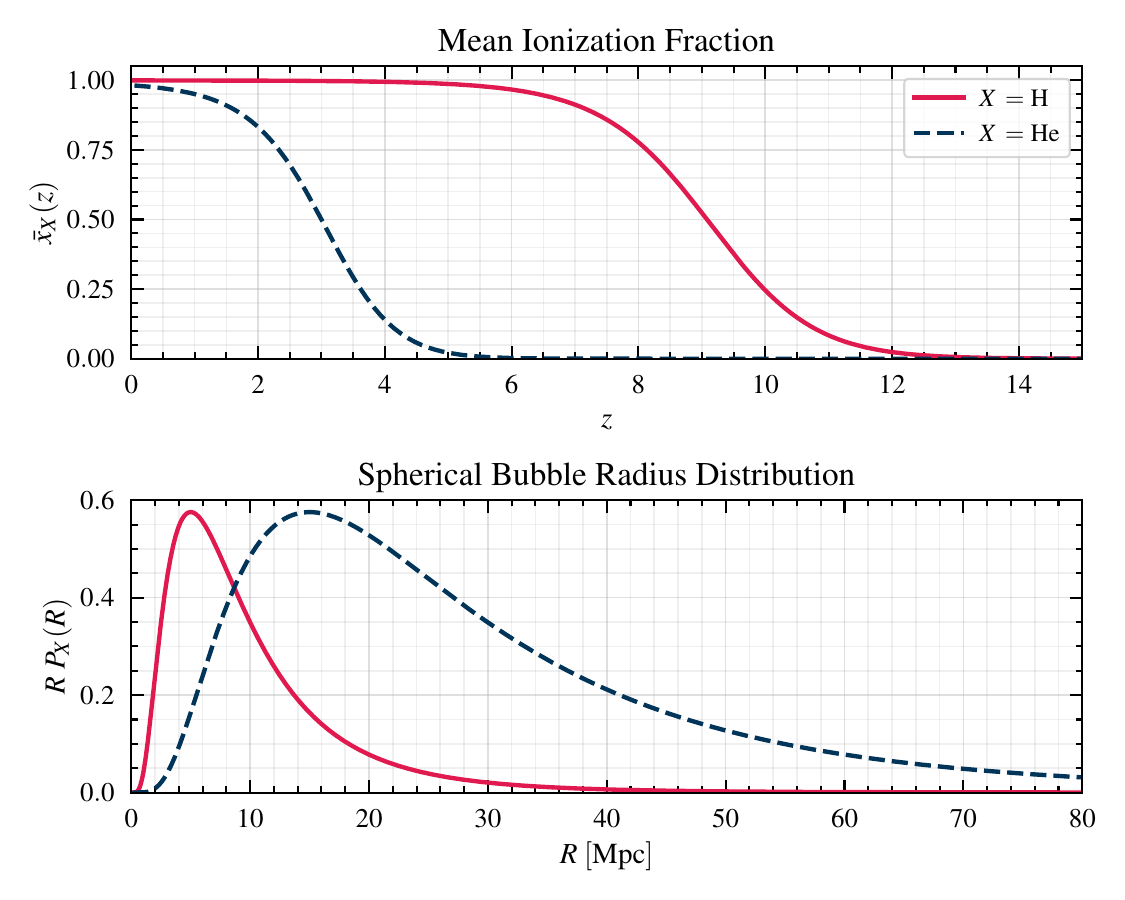}
    \caption{Summary of the patchy-reionization model assumed in our forecasts.
    \textit{Upper panel:} the assumed redshift-evolution of the mean ionization fraction $\bar{x}_X$ for both $X =$ H and $X =$ He.
    \textit{Lower panel:} spherical bubble-radius distribution $R\,P_X(R)$ for both the epochs. The functional forms corresponding to the plots in the upper and lower panel are given by Eqs.~\eqref{eq:mean_ionization_fraction_functional_form} \& \eqref{eq:bubble_radius_distribution_functional_form}, respectively. The assumed fiducial parameters can be found in Tab.~\ref{tab:set_of_params}. 
    }
    \label{fig:reionization_simplified_models}
\end{figure}

\begin{table}
    \centering
    \caption{\textit{Fiducial values of reionization parameters for H and He:} Summary of the fiducial values assumed to model the epochs of patchy-H and -He reionization. 
    The first pair of parameters $\{y_{\rm re}^{X},\, \Delta_y^{X}\}$ govern the redshift evolution of $\bar{x}_X(z)$ [Eq.~\eqref{eq:mean_ionization_fraction_functional_form}] and second pair $\{\bar{R}^{X},\, \sigma_{\ln{R}}^{X}\}$ determine the ionized-bubble-radius distribution $P_X(R)$ [Eq.~\eqref{eq:bubble_radius_distribution_functional_form}].
    }
    \label{tab:set_of_params}
    \vspace{10pt}
    \renewcommand{\arraystretch}{1.4} 
    \setlength{\tabcolsep}{12pt}     
    \centering
    \small
    \begin{tabular}{c|c c}
    \hline \hline
    Parameter & \multicolumn{2}{c}{Fiducial Value}  \\ \cline{2-3}
              & \ Hydrogen \ & Helium \ \\ \hline
    $y_{\rm re}^{X}$                      & $33.5$        & $8.0$     \\
    $\Delta_y^{X}$                        & $8.0$         & $3.5$      \\
    $\bar{R}^{X}$                         & $5\ \rm{Mpc}$ & $15\ \rm{Mpc} \ \ $\\
    $\sigma_{\ln{R}}^{X}$                 & $\ln 2.0$     & $\ln 2.0$ \\
    Bubble Bias $b^{X}$          & 6.0           & 6.0     \\ \hline \hline
    \end{tabular}
\end{table}

Figure~\ref{fig:reionization_simplified_models} displays the fiducial models assumed for the epochs of H and He reionization. The top panel displays the redshift evolution of the average ionization fraction and the bottom panel displays the fiducial, ionized bubble-radius distribution. 
In each panel, the pink-solid and blue-dashed curve correspond to the model adopted for H and He reionization, respectively.
In addition to the fiducial-parameter values listed in this section, we also assume that the bubble bias $b^{X} = 6.0$ for both $X \in \{{\rm H},\, {\rm He}\}$. 
Although we do not consider alternate models with varied values of this parameter, final expressions for the $P_{ee}^{\rm ion}(k,z)$ in the following section demonstrate that an increase in this parameter value simply results in an increase in power from patchy-reionization on all scales, likely leading to tighter constraints on reionization model parameters. 
The set of fiducial parameters assumed for the signals and forecasts presented in Sec.~\ref{sec: kSZ Trispectrum and Galaxy Cross-Correlation} and Sec.~\ref{sec:Forecasts} are summarized in Tab~\ref{tab:set_of_params}. 
Although this simple, semi-analytical model does not contain some of the qualitative features observed in simulations of both the epochs of reionization, such as the dependence of $P_X(R)$ on redshift~\cite{Zahn:2006sg} or the dependence of $b^X$ on the size of ionized regions, the restricted parameter space allows us to get a qualitative sense of the information contained in the proposed cross-correlation probes and compare our results more easily with other works that have adopted similar models~\cite{Dvorkin:2008tf, Caliskan:2023yov}. 

\subsection{Ionized Electron Power Spectrum}
\begin{figure*}
    \centering
    \includegraphics[width=\linewidth]{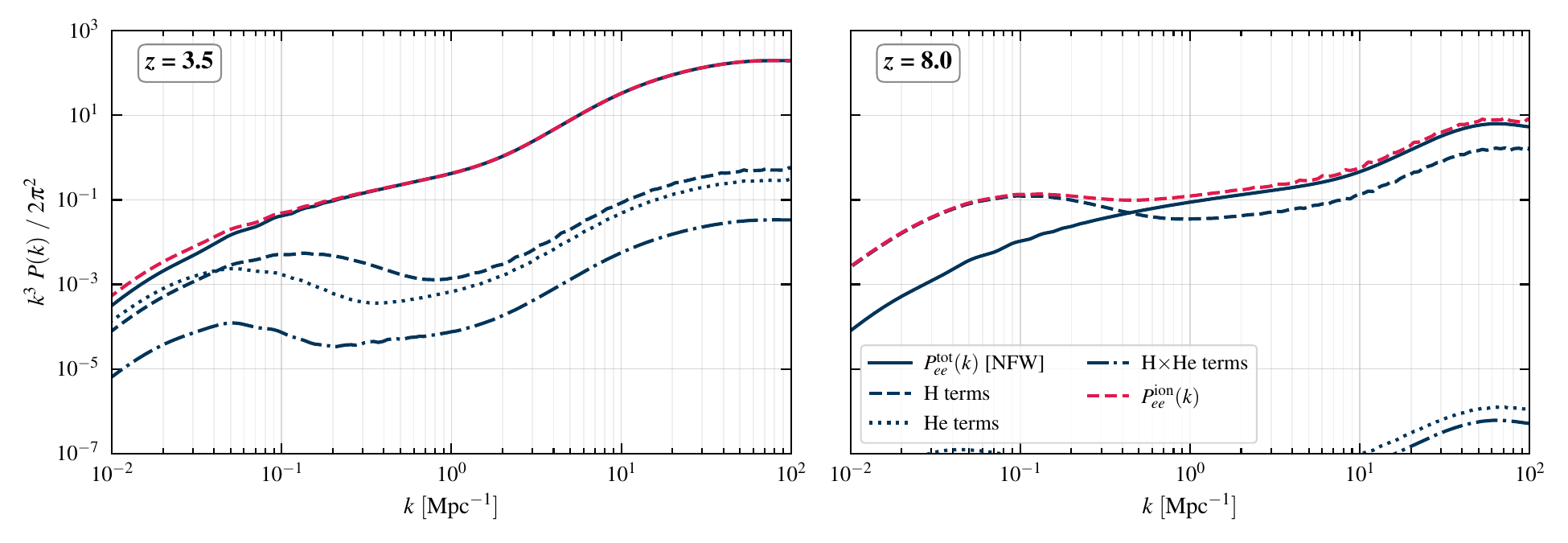}
    \caption{\textit{Left:} the pink-dashed curve displays ionized electron power spectrum $P_{ee}^{\rm ion}(k,z)$ [calculated using Eq.~\eqref{eq: ion_elec_PS_as_ft_2pt_terms}, assuming an NFW profile for $u_e(k|M,z)$] at redshift $z=3.5$, alongside its separate contributors in the set of blue lines. The solid-blue curve is the contribution from the first term in Eq.~\eqref{eq: ion_elec_PS_as_ft_2pt_terms}, sourced solely by fluctuations in $\netot(\br, z)$. The blue-dashed [blue-dotted] curve is the contribution from terms that depend on $\xH(\br, z)$ [$\xHe(\br, z)$]. These may include a dependence on $\netot(\br,z)$, but do not include terms that are dependent on \textit{both} $\xHe(\br,z)$ and $\xH(\br,z)$. The blue dot-dashed curve represents the contributions from `mixed' terms that depend on both H and He reionization. \textit{Right:} the same quantities as the left panel, but for  $z = 8.0$.
    }
    \label{fig:ionized_elec_PS_all_parts_two_redshifts}
\end{figure*}
With models for $\delta\netot(\bk,z)$, $\delta\xH(\bk,z)$, and $\delta\xHe(\bk,z)$ set up, we can finally compute the terms in Eq.~\eqref{eq: ion_elec_two_point_terms_real_space} to obtain the ionized electron power spectrum [Eq.~\eqref{eq: ion_elec_PS_as_ft}]. 
Given Eqs.~\eqref{eq: ion_elec_two_point_terms_real_space} and \eqref{eq: ion_elec_PS_as_ft}, we have:
\begin{widetext}
    \begin{eqnarray}
        P_{ee}^{\rm ion}(k) = \frac{\VEV{\netot\netot}_k}{(\bnetot)^2} 
        + \frac{\VEV{\xtot\xtot}_k }{(\bxtot)^2} 
        + \frac{2\VEV{\netot\xtot}_k}{\bxtot\bnetot}  + \frac{\lr{[}{\left(\VEV{\xtot\xtot}\star\VEV{\netot\netot}\right)(k) + \left(\VEV{\netot\xtot}\star\VEV{\netot\xtot}\right)(k)}{]}}{(\bar{n}_e^{\rm ion})^2}\,,
        \label{eq: ion_elec_PS_as_ft_2pt_terms}
    \end{eqnarray}
\end{widetext}
where the redshift dependence of $P_{ee}(k, z)$, $\bnetot$, $\bxtot$, and all the correlation functions of the form $\VEV{\dots}_k$ has been suppressed for ease of notation. 
It is important to note that in the above expression for $P_{ee}^{\rm ion}(k,z)$ we have assumed that the four-point function $\VEV{(\delta\xtot\cdot\delta\netot)(\delta\xtot\delta\netot)}_r$  can be expanded into the terms $\VEV{\xtot\xtot}_r\VEV{\netot\netot}_r - \VEV{\netot\xtot}_r^2 $ using Wick's theorem. These products of correlation functions in real-space are then represented as convolutions ($\star$-operator) in Fourier-space. 

Since $\delta\xtot(\bk, z)$ gets a contribution from $\delta\xH(\bk, z)$ and $\delta\xHe(\bk, z)$ [Eq.~\eqref{eq: ionized_elec_base_model}], we need to calculate Fourier-space two-point functions of the form $\VEV{x_Xx_{X'}}_k$ and $\VEV{x_X\netot}_k$ for each $X, X' \in \{{\rm H},\, {\rm He}\}$, as well as $\VEV{\netot\netot}_k$. 
Given the HOD expression for $\delta\netot(\bk, z)$ in Eq.~\eqref{eq: tot_elec_dist_fourier}, and the HOD-based model for $\delta x_X(\bk,z)$ in Eq.~\eqref{eq: x_X_fourier_distrib_simpl}, these correlation functions take the following form:
\begin{widetext}
    \begin{eqnarray}
        \VEV{\netot\netot}_k &= &(\bnetot)^2\beta_e(k)^2\Plin(k) + \lr{[}{\int \dd M\,n(M)N_e(M)u_e(k|M)}{]}^2 \,, \label{eq: tot_elec_dist_fourier_2pt}\\
        \VEV{\netot x_X}_k &= &\bnetot\bar{x}_X\beta_e(k)\beta_X(k)\Plin(k) + \int \dd M\dd R\, n(M) N_e(M)V_b(R)P_X(R)u_e(k|M)W_R(k)\,, \label{eq: x_X_fourier_simple_two_point}\\
        \VEV{x_X x_{X'}}_k &= &\bar{x}_X\bar{x}_{X'}\beta_X(k)\beta_{X'}(k)\Plin(k) + \frac{\delta_{XX'}}{\VEV{V_b}}(\bar{x}_X - \bar{x}_X^2)\int \dd R\, P_X(R)V_b(R)^2W_R(k)^2\,, \label{eq: ne_tot_x_X_fourier_two_point}
    \end{eqnarray}
\end{widetext}
where $X \in \{{\rm H},\, {\rm He}\}$, and we have suppressed the redshift dependence of some quantities for ease of notation. The terms $\beta_e(k,z)$ and $\beta_X(k,z)$, used in the above equation, are defined as:
\begin{align}
    \beta_e(k) &\equiv 1 + \int \frac{\dd M}{\rho_m} M n(M) b(M) \left[ u_e(k|M) - 1 \right], \\
    \beta_X(k) &\equiv \frac{[1 - \bar{x}_X] \ln[1 - \bar{x}_X]}{\bar{x}_X^2 \VEV{V_b}} \nonumber \\
    &\quad\hspace{3.75em}\times b^X \int \dd R \, P_X(R) V_b(R) W_R(k)\,.
\end{align}
The two point functions presented in Eqs.~\eqref{eq: tot_elec_dist_fourier_2pt}-\eqref{eq: ne_tot_x_X_fourier_two_point} culminate in an ionized electron power spectrum that is very \textit{similar} to what other forecast papers have used in the past (see, for example, Refs.~\cite{Dvorkin:2008tf, Caliskan:2023yov}), with the result for both the 1$h$- and $2h$-terms of $\VEV{x_Hx_H}_k$ matching the expressions derived in Ref.~\cite{Mortonson:2006re} exactly. 
However, it is important to note some key differences. 
In our model for $P_{ee}^{\rm ion}(k,z)$ we have intentionally introduced a dependence on the small-scale distribution of total electrons. 
This allows us to make informed forecasts about how the small-scale power in $\VEV{\netot\netot}_k$ impacts our projected constraints.  
For consistency, this dependence on $u_e(k|M,z)$ is also incorporated into our expression for $\VEV{\netot x_X}_k$ as well.
Finally, our model for $P_{ee}^{\rm ion}(k,z)$ also includes a cross-correlation term between spatial fluctuations in $\xH(\br, z)$ and $\xHe(\br, z)$, since our model for $\xtot$ allows for a period of overlap between the two epochs. 
However, we choose to ignore a $1h$-like term in $\VEV{\xH\xHe}_k$ for the forecasts in this work. 

Figure~\ref{fig:ionized_elec_PS_all_parts_two_redshifts} displays $P_{ee}^{\rm ion}(k, z)$ at redshifts $z=3.5$ (left) and $z=8.0$ (right) in the pink-dashed curves. 
The former redshift ($z=3.5$) corresponds to a time at which our model predicts significant patchiness in the (second) ionization of neutral He. 
In contrast, the latter one displays the same power spectrum when a significant amount of total power is sourced by fluctuations in $\xH(\br, z)$. 
The figure also displays the separate contributions from $\netot(\br, z)$ and $x_X(\br, z)$ (for $X\in\{\mathrm{H},\,\mathrm{He}\}$) to $P_{ee}^\mathrm{ion}(k)$ via the set of dark-blue curves. 
The solid-blue curve, labeled $P_{ee}^{\rm tot}(k)$, displays the contribution from the first term in Eq.~\eqref{eq: ion_elec_PS_as_ft_2pt_terms}, assuming an NFW profile for $u_e(k|M,z)$. 
In other words, it represents the contribution from $P_{ee}^{\rm tot}(k,z) \equiv \VEV{\netot\netot}_k/ (\bnetot)^2$. 
In contrast, dashed-blue [dotted-blue] curve, labeled `H-terms' [`He-terms'],  represents the contribution to $P_{ee}^{\rm ion}(k,z)$ coming \textit{solely} from fluctuations in the $\xH(\br,z)$ [$\xHe(\br, z)$] field. 
This includes terms that are (also) dependent on $\netot(\br, z)$ but not terms that are sourced by \textit{both} $\xH(\br,z)$ and $\xHe(\br, z)$. 
The contributions from these `mixed' terms that depend on reionization model parameters across both epochs are included as the dot-dashed blue curve. 
As expected, the contribution from `H-terms' is much more significant at $z=8.0$, surpassing the power in $P_{ee}^{\rm tot}(k)$ on large-scales. Similarly, the contribution from `He-terms' is visibly higher at $z=3.5$, although the power remains bounded by the contributions from `H-terms' and $P_{ee}^{\rm tot}$.

\section{kSZ Trispectrum and Galaxy Cross-Correlation}
\label{sec: kSZ Trispectrum and Galaxy Cross-Correlation}
In this section, we provide a summary of the four-point statistic (trispectrum) that not only isolates the kSZ effect but also leverages the dependence of this CMB secondary on the cosmological velocity field in order to give more information about its source redshift dependence. 
It was initially introduced in Ref.~\cite{Smith:2016lnt}, within the context of characterizing the epoch of H reionization, as a complementary probe to 21-cm measurements~\citep[e.g.,][]{Liu:2019awk,Park:2018ljd,Weltman:2018zrl,Patil:2017zqk,Beardsley:2016njr,DeBoer:2016tnn,Silva:2014ira}.
In the following sub-sections, we present an extension of existing analysis by explicitly introducing the dependence of this trispectrum on He reionization, given the model presented Sec.~\ref{sec: Reionization Model for the Ionized Electron Power Spectrum}. 

Subsequently, we introduce our model for the cross-correlation trispectrum, constructed from cross-correlating the kSZ temperature-squared map with the large-scale radial velocity field (squared) obtained from galaxy-survey data. Given a prescription for the trispectrum signal, with a clear dependence on our analytical models for reionization above, we display the expected contributions from each of the epochs of reionization to the signal constructed from kSZ maps alone. Furthermore, to establish the advantage of cross-correlation, we conclude with a demonstration of the expected contributions from both H and He reionization to the proposed cross-spectrum. 

\subsection{The kSZ Trispectrum}
\label{subsec: The kSZ Trispectrum}
The temperature fluctuations in the observed CMB data, sourced by the kSZ effect, can be expressed as:
\begin{eqnarray}
    \frac{\Delta T_{\rm kSZ}(\bn)}{T_{\rm CMB}} = \int \frac{\dd \chi\,\sigma_T}{(1+z)^{-2}}e^{-\bar{\tau}(z)}\neion(\bn, z)\bn\cdot \bv(\bn, z)\,,\nonumber\\ 
\end{eqnarray}
where $T_{\rm CMB}$ is the average CMB temperature, $\chi(z)$ represents comoving distance to redshift $z$, $\sigma_T$ is the Thomson scattering cross-section, $\bar{\tau}(z)$ is the average optical depth of scattering up to redshift $z$, and  $\bv(\bn, z)$ represents the cosmological bulk velocity at sky-location $\bn$ and redshift $z$. 
In this work, we approximate the power spectrum of the kSZ temperature anisotropy using the following integral~\cite{Ma:2001xr}:
\begin{eqnarray}
    C_\ell^{\rm kSZ} = \int \dd z Q(z)\VEV{v_r(z)^2}P_{ee}^{\rm ion}\lr{(}{\frac{\ell}{\chi(z)}, z}{)}\,,
\end{eqnarray}
where $\VEV{v_r^2(z)} = \VEV{v^2(z)}/3$ is the mean-squared radial velocity at redshift $z$, and the radial weight function $Q(z)$ is defined as:
\begin{eqnarray}
    Q(z) \equiv T_{\rm CMB}^2 \frac{H(z)}{\chi(z)^2}\lr{[}{\frac{\dd \bar{\tau}(z)}{\dd z}}{]}^2e^{-2\bar{\tau}(z)}\,.
\end{eqnarray}

\begin{figure}
    \centering
    \includegraphics[width=0.5\textwidth]{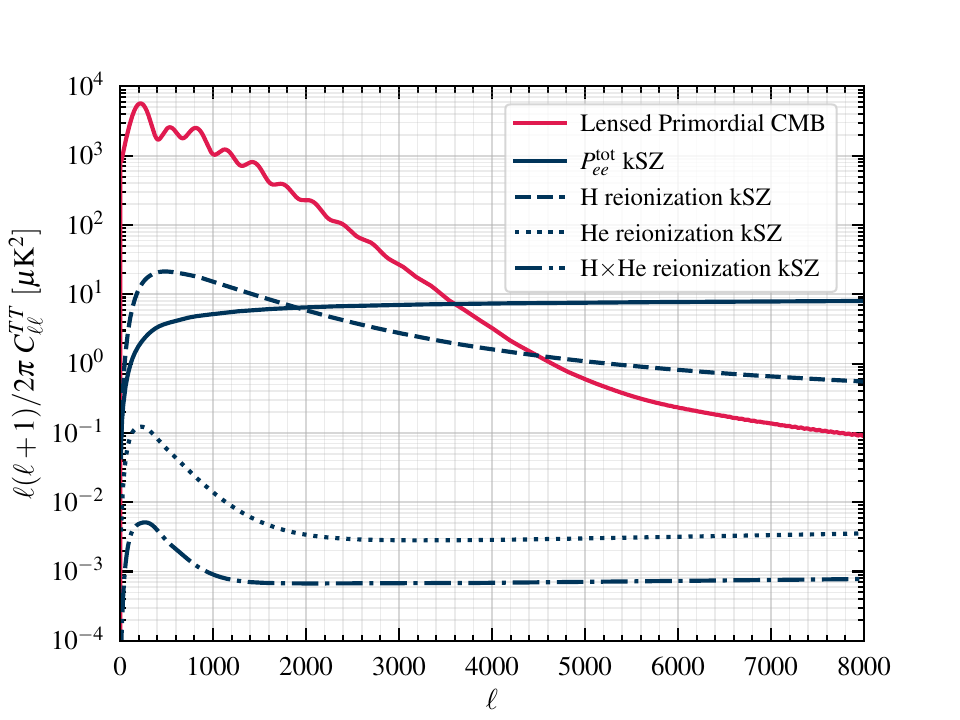}
    \caption{
    Lensed-primary CMB (pink-solid curve) and separate contributions to $C_\ell^{\rm kSZ}$ (blue curves). The blue-solid curve represents the redshift-integrated contribution from $P_{ee}^{\rm tot}(k,z)$ [first term in Eq.~\eqref{eq: ion_elec_PS_as_ft_2pt_terms}]. The dashed-blue (dotted-blue) curve is the contribution from terms that are dependent on H (He) reionization model-parameters [possibly including a dependence on $\netot(\br, z)$]. The contribution from the remaining terms that depend on \textit{both} $\xH(\br,z)$ and $\xHe(\br,z)$ is displayed as the blue dot-dashed curve. Signals are calculated assuming an NFW (AGN) profile for $u_e(k|M,z)$ at $z \gtrsim 5$ ($z \lesssim 5$).}
    \label{fig:kSZ_and_prim_CMB_all_parts}
\end{figure}

As a result, the morphology of patchy H and He reionization is imprinted on the observed kSZ signal via its effect on $P_{ee}^{\rm ion}(k,z)$ [Eq.~\eqref{eq: ion_elec_PS_as_ft_2pt_terms}]. Figure~\ref{fig:kSZ_and_prim_CMB_all_parts} displays the separate contributions to the total integrated kSZ effect. 
As done previously with Fig.~\ref{fig:ionized_elec_PS_all_parts_two_redshifts}, the solid-blue curve represents the contribution to $C_\ell^{\rm kSZ}$ from $P_{ee}^{\rm tot}(k,z)$ alone, integrated along the line of sight. 
Similarly, the dashed-blue [dotted-blue] curve represents the contribution from terms in Eq.~\eqref{eq: ion_elec_PS_as_ft_2pt_terms} that are dependent only on the H-ionization [He-ionization] field or some combination of $\xH(\br, z)$ [$\xHe(\br,z)$] and $\netot(\br, z)$. 
Finally, the redshift-integrated contribution from terms that are dependent on \textit{both} $\xH(\br, z)$ and $\xHe(\br,z)$ is represented by the blue-dot-dashed curve. 
The lensed, primordial CMB signal is plotted as the pink-solid curve in the same frame for comparison.  

The curves plotted in Fig.~\ref{fig:kSZ_and_prim_CMB_all_parts} indicate that on small scales, where the kSZ effect is most dominant, the power from each of the epochs of reionization does not have a unique $\ell$-dependence that can be clearly distinguished from the contribution of $P_{ee}^{\rm tot}(k,z)$. 
In other words, using the kSZ power spectrum alone to probe the epochs of reionization will likely be a difficult task. 
Therefore, to optimally leverage the kSZ effect we turn to the higher-order statistic explored in Refs.~\cite{Smith:2016lnt, Ferraro:2018izc}.

To convert the expression for kSZ anisotropies into the approximate expression for the trispectrum from Refs.~\cite{Smith:2016lnt, Ferraro:2018izc}, we start by defining a set of band-limited, high-pass filters $\{W_{S, i}(\ell)\}$ for $i \in [1, \Nellbins]$ as follows:
\begin{equation}
    W_{S,i}(\ell) = \begin{cases}
        (C_\ell^{\rm kSZ})^{1/2}/\tilde{C}_{\ell}^{\rm tot}&  \textrm{if }\ell_{\rm min}^i \leq \ell \leq \ell_{\rm max}^i\,, \\
        0& \textrm{otherwise}\,,
    \end{cases}
\end{equation}
where $\ell_{\rm min}^i$ and $\ell_{\rm min}^i$ bound the $i^{\rm th}$ $\ell$-band. With each of the above set of filters applied to the harmonic-space CMB temperature map as $T_{S,i}(\ell) = W_{S,i}(\ell)T(\ell)$, one can define a set of $\Nellbins$ fields 
$\{K_i(\bn)\}$, where each field $K_i(\bn)$ is obtained by squaring $T_{S,i}(\bn)$ in real space. 
Physically, each field $K_i(\bn)$ then represents the locally measured, small-scale power in direction $\bn$ for $\ell_{\rm min}^i \leq \ell \leq \ell_{\rm max}^i$. 
Therefore, the sky-averaged small-scale power in the $i^{\rm th}$ high-$\ell$ band can be written as:
\begin{eqnarray}
    \bar{K}_{{\rm tot}, i} = \int \frac{\dd^2 \bl}{(2\pi)^2}W_{S, i}(\ell)^2\tilde{C}_{\ell}^{\rm tot}\,,
\end{eqnarray}
where $\tilde{C}_{\ell}^{\rm tot}$ is the observed CMB signal, comprised of the lensed primary CMB signal $C_\ell^{TT}$, the kSZ effect $C_\ell^{\rm kSZ}$, as well as contamination from instrument noise, and frequency-dependent foregrounds post-ILC cleaning (detailed in Sec.~\ref{subsec: Experiment Specifications}). 

Assuming that the non-Gaussian contribution to the field $K_i(\bn)$ is sourced by the kSZ effect, the contribution to $\bar{K}_i$ from each redshift bin can be written as:
\begin{eqnarray}
    \frac{\dd\bar{K}_i}{\dd z} = \int \frac{\dd^2\bl}{(2\pi)^2} W_{S,i}^2(\ell)\frac{\dd C_\ell^{\rm kSZ}}{\dd z}.
\end{eqnarray}
At this point, we note that for a fixed realization of the radial velocity field $v_r(\bn,z)$, large-scale modulations in the kSZ effect at redshift $z$ can be attributed to electron distributions (`patchy' only on small-scales) experiencing long-wavelength (cosmological) radial-velocity perturbations. 
Then, on large angular scales, each field $K_{i}(\bn)$ can be approximately modeled as:
\begin{eqnarray}
    K_i(\bn) = \int \dd z \frac{\dd \bar{K}_i}{\dd z} \eta(\bn, z)\,
    \label{eq: K_nhat_eta_model}
\end{eqnarray}
where $\eta(\bn, z) = v_r(\bn, z)^2/\VEV{v_r^2(z)}$. 

The above construction indicates that there are two separate scales to be considered in the binned-trispectrum statistic. 
Most prominently, this statistic leverages the fact that large-scale modulations in $K_i(\bn)$ are sourced by long-wavelength perturbations in $v_r(\bn,z)^2$ along the line-of-sight. 
Since the velocity-field has a well-defined coherence length, it acts as a ``standard-ruler'', i.e, $v_r(\bn,z)^2$ in different redshift bins induces correlations in the $C_L^{K_iK_j}$ power spectrum (trispectrum-statistic) on different angular scales. 
Consequently, the $ L $-dependence of $ C_{K_iK_j}^{L} $ facilitates clearer differentiation between epochs separated along the line of sight. 
Additionally, grouping small-scale modes ($ \ell \gtrsim 3000 $) into distinct bins, to construct the suite of fields  $\{K_i\}$ for $i \in [1,\, \Nellbins]$, preserves the valuable scale-dependent information contained in each individual component of the kSZ effect, thereby enhancing the sensitivity and interpretability of the resulting forecasts.

With this motivation in hand, under the Limber approximation, $C_L^{K_iK_j}$ can be written as:
\begin{eqnarray}
    C_L^{K_iK_j} = \int \dd z \frac{H(z)}{\chi(z)^2}\frac{\dd \bar{K}_i}{dz}\frac{\dd \bar{K}_j}{dz}P_{\eta\eta}^\perp(k = L/\chi, z)\,.
    \label{eq: C_L_KK_model_expression}
\end{eqnarray}
In the above equation, $P_{\eta\eta}^\perp(k,z)$ refers to the power spectrum of the $\eta(\bn, z)$ field, evaluated at wavenumber $\bk$ perpendicular to the line of sight. 
We calculate this power-spectrum in the linear-theory limit:
\begin{equation}
\begin{split}
P_{\eta\eta}^{\perp}(k)\!=\!\frac{2}{\langle v_r(z)^2\rangle^2}\!\!\int\!\!&\frac{\dd^3\boldsymbol{k}'}{(2\pi)^3}\frac{(k_r')^2(k_r-k_r')^2}{(k')^2\,(|\bk-\bk'|)^2}\\
&\times\!P_{vv}(k', z)P_{vv}(|\bk-\bk'|,z)\,,
\end{split}
\label{eq: p_eta_perp}
\end{equation}
where $P_{vv}(k,z)$ velocity power-spectrum, computed in the linear regime and the above integral is be computed assuming $k_r = 0$. 
Although the integrand in the above equation is redshift-dependent, in the linear regime the above expression for $P_{\eta\eta}^{\perp}(k)$ results in a $z$-independent quantity. 
Further details on the calculation of the above integral can be found in Appendix~\ref{appendix: Calculation of P_eta Integral}.

Given the power spectrum in Eq.~\eqref{eq: C_L_KK_model_expression}, which quantifies the clustering power induced by correlations in the velocity field sourcing the kSZ effect, the noise is then the value that $C_L^{K_iK_j}$ would take if the small-scale temperature field was purely Gaussian. 
This results in the following expression for reconstruction noise $N_L^{K_iK_j}$:
\begin{eqnarray}
    N_L^{K_iK_j} = 2\delta_{ij}\int \frac{\dd^2\bl}{(2\pi)^2}W^2_{S, i}(\ell)W^2_{S,i}(|\boldsymbol{L}-\bl|)C_\ell^{\rm tot}C_{|\boldsymbol{L} - \bl|}^{\rm tot}\,,\nonumber\\
    \label{eq: N_L_KK_expression}
\end{eqnarray}
where the Kronecker delta function $\delta_{ij}$ ensures that $N_L^{K_iK_j} = 0$ for $i\neq j$. 
It is important to note that this is the expected noise in the kSZ-sourced $C_L^{K_iK_j}$, assuming no additional non-Gaussian contributions to the modeled signal (from CMB lensing or the cosmic infrared background, for example).
From this point on, when referring to the suite of angular power-spectra $\{C_L^{K_iK_j}, N_L^{K_iK_j}\}$ for $i,j \in [1, \Nellbins]$, we will use shorthand $C_L^{KK}$ (or simply $KK$) and $N_L^{KK}$, dropping the lowered indices.

\begin{figure}[h!]
    \centering
    \includegraphics[width=\linewidth]{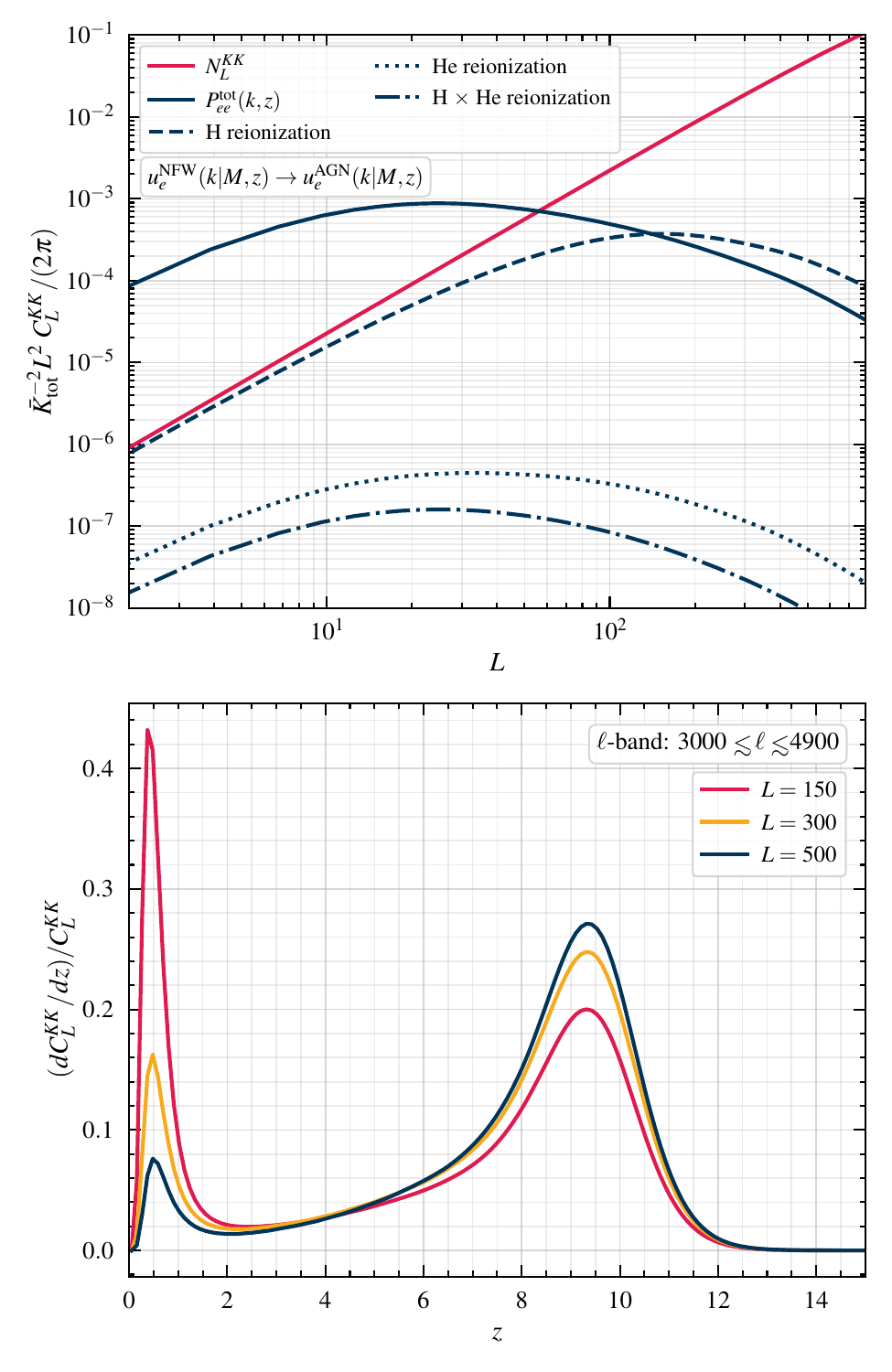}
    \caption{\textit{Top:} separate contributions to the integrated $C_L^{KK}$ signal (blue) and noise $N_L^{KK}$ (pink). The solid blue curve represents the contribution to $C_L^{KK}$ sourced \textit{solely} by fluctuations in $\netot(\br,z)$. The dashed-blue [dotted-blue] curve labeled `H reionization' [`He-reionization'] corresponds to the contribution to $C_L^{KK}$ sourced by fluctuations in $\xH(\br,z)$ [$\xHe(\br,z)$]. This includes terms from Eq.~\eqref{eq: ion_elec_PS_as_ft_2pt_terms} [propagated to Eq.~\eqref{eq: C_L_KK_model_expression}] that may be dependent on $\netot(\br, z)$ but are not dependent on fluctuations in \textit{both} $\xH(\br,z)$ and $\xHe(\br,z)$. The contribution from these mixed terms is represented by the dot-dashed blue curve labeled `H$\times$He reionization'. \textit{Bottom:} contributions to $C_L^{KK}$ as a function of source redshift $z$ for different values of $L$. All the presented curves have been calculated for $3000 \lesssim \ell \lesssim 4900$ [lowest $\ell$-band in CMB-HD-based forecasts], assuming that $C_\ell^{\rm tot} = C_\ell^{TT} + C_\ell^{\rm kSZ} + N_\ell^{\rm HD} + F_\ell^{\rm HD}$, where $N_\ell^{\rm HD}$ and $F_\ell^{\rm HD}$ account for instrument noise and foregrounds expected from a CMB-HD-like survey (for details, see Sec~\ref{subsec: Experiment Specifications}). Signals are calculated using the NFW (AGN) profile for $u_e(k|M,z)$ at $z \gtrsim 5$ ($z \lesssim 5$).
    }
    \label{fig:KK_signal_and_noise_all_parts_with_dz}
\end{figure}

For the forecasts presented in Sec~\ref{sec:Forecasts}, the set of filters $W_{S,i}(\ell)$ are constructed for equal-width $\ell$-bands, with the lowest band beginning at $3000$ and the highest band ending at $\ell_{\rm max}$, the smallest accessible scale for the CMB survey in consideration. 
An example of the $C_L^{K_iK_i}$ power spectrum, along with its source-redshift dependence, is presented in Fig.~\ref{fig:KK_signal_and_noise_all_parts_with_dz}. 
All the presented curves have been calculated for the lowest $\ell$-band considered in the CMB-HD forecasts ($3000 \lesssim \ell  \lesssim 4900$). The weights $W_{S,1}(\ell)$ are computed assuming that $\tilde{C}_\ell^{\rm tot} = C_\ell^{TT} + C_\ell^{\rm kSZ} + N_\ell^{\rm HD} + F_\ell^{\rm HD}$, where $N_\ell^{\rm HD}$ and $F_\ell^{\rm HD}$ account for contamination in the observed signal from instrument noise and foregrounds, computed assuming survey specifications matching CMB-HD (Sec~\ref{subsec: Experiment Specifications}).

The top panel of Fig.~\ref{fig:KK_signal_and_noise_all_parts_with_dz} displays the separate contributions to the integrated $C_L^{K_1K_1}$ signal. 
These signals have been calculated assuming that electron distributions are characterized by the NFW (AGN) profile for $z \gtrsim 5$ ($z \lesssim 5$). 
Consistent with previous figures, the solid blue curve corresponds to the contribution to $C_L^{K_1K_1}$ sourced \textit{only} by the clustering properties of $\netot(\br,z)$ [first term in Eq.~\eqref{eq: ion_elec_PS_as_ft_2pt_terms}, propagated through to Eq.~\eqref{eq: C_L_KK_model_expression}]. 
In contrast, the dashed-blue [dotted-blue] curve accounts for anisotropies in $K_1(\bn)$ sourced by fluctuations in $\xH(\br,z)$ [$\xHe(\br,z)$]. These terms may include a dependence on $\netot(\br,z)$. 
The contribution from the `mixed' terms that are dependent on characteristics of patchy-reionization across both epochs is depicted via the blue dot-dashed curve. 
The pink curve presents an example of $N_L^{K_1K_1}$ for comparison to the expected signals.

An important property of the $C_L^{KK}$ signal is that the contribution from redshift $z$ is proportional to $P_{\eta\eta}^{\perp}(k = L/\chi)$, with no additional $L$-dependence. 
However, the amplitude of this contribution is set by the small-scale physics of $\dd\bar{K}/\dd z$. 
The bottom panel of Fig.~\ref{fig:KK_signal_and_noise_all_parts_with_dz} displays the source redshift dependence of $C_L^{K_1K_1}$ for different values of $L$. 
The separate curves in the plot suggest that smaller values of $L$ will contain significantly more power sourced by small-scale physics at late times. 
This is consistent with the expected projections of the $v_r(\bn, z)$ coherence length at lower redshifts on the two-dimensional surface of the CMB sky. 

At this point, it is important to note a few key differences between the final form of $C_L^{KK}$ used in the forecasts below and the original model of the signal proposed in Ref.~\cite{Smith:2016lnt}.
Comparing Fig.~\ref{fig:KK_signal_and_noise_all_parts_with_dz} with the seemingly analogous Fig. 2 of Ref.~\cite{Smith:2016lnt} reveals that the `late-time' contribution to $C_L^{KK}$ is much stronger than the contribution from H reionization, contrary to the signals presented in Ref,~\citep{Smith:2016lnt}.
The main reason behind this discrepancy is the difference in our chosen model for $P_{ee}^{\rm ion}(k,z)$. 
First, we find that decreasing small-scale power in $P_{ee}^{\rm tot}$ [see Fig.~\ref{fig:total_elec_PS_no_re} for possible scenarios] results in a visibly lower contribution from fluctuations in $\netot(\br,z)$ to the observed $C_L^{KK}$ signal.
In other words, assuming that $u_e(k|M,z)$ follows the $W_e(k)\times$NFW profile at $z\lesssim 5$, as opposed to the assumed AGN distribution, would result in a diminished overall amplitude of the $P_{ee}^{\rm tot}(k)$ curve. 
Moreover, we find that our chosen model for $\xH(\br, z)$ predicts noticeably smaller power from H reionization at small-scales in the kSZ effect (Fig.~\ref{fig:kSZ_and_prim_CMB_all_parts}), resulting in its relatively suppressed contribution to $C_L^{KK}$. 
The H-sourced kSZ power on small scales may be boosted by a longer duration of reionization, such that the photons encounter more anisotropy along the line of sight. 
Conversely, power at high-$\ell$ can also be boosted by assuming that the average size of ionized regions during the epoch is smaller. 
Therefore, reasonable changes to $P_{ee}^{\rm tot}$ and $\xH(\br, z)$ will present results more compatible with Ref.~\cite{Smith:2016lnt}.
 
The results in Fig.~\ref{fig:KK_signal_and_noise_all_parts_with_dz} bring to light two noteworthy observations. On the one hand, the comparable power across the contributions from H reionization and $P_{ee}^{\rm tot}(k,z)$, alongside the distinctly double-peaked nature of $\dd C_L^{K_iK_j}/\dd z$, indicates that the trispectrum signal $C_L^{KK}$ has the potential to constrain the patchy morphology of H reionization. 
Contrarily, the comparatively lower power in the contributions from any He-dependent terms to $C_L^{KK}$, alongside the lack of any visible features in the source redshift dependence of $C_L^{KK}$ near $z\sim 3$, demonstrates that using the redshift-integrated probe $K_i(\bn)$ may not allow for the reconstruction of patchy He reionization. 
However, if we are able to construct a cross-correlation statistic that can supplement $K(\bn)$ at low redshifts, or possibly trace the redshift evolution of $P_{ee}^{\rm ion}$ for $z \lesssim 5$, we may be able to constrain both epochs of reionization simultaneously. 
To this end, the galaxy-survey data set offers itself as a promising probe.

\subsection{The galaxy-kSZ cross correlation}
\label{subsec:The galaxy-kSZ cross correlation}
Given the continuity-equation-based relation between the cosmological velocity field $\boldsymbol{v}(\bk, z)$ and the large-scale matter over-density field $\delta_m(\bk, z)$, measurements of galaxy distributions from large-scale galaxy surveys can be used to re-construct long-wavelength perturbations in $v(\bk, z)$ as follows:
\begin{eqnarray}
    \hatv(\bk,z)= \frac{f(z)H(z)}{k(1+z)}\frac{b_v(z)}{b_g(z)}{\hat{\delta}_g(\boldsymbol{k},z)}\,.
    \label{eq: galaxy_density_to_radial_velocity}
\end{eqnarray}
In the above equation, $\hatv(\bk,z)$ is the Fourier domain, large-scale velocity field specifically reconstructed from galaxy-survey data, $f(z)$ refers to the linear growth rate $\dd \ln G/\dd \ln a$, $\hat{\delta}_g(\bk,z)$ is the \textit{observed} galaxy density field, and $b_g(z)$ is the large-scale galaxy bias [i.e., $P_{gg}(k,z) \equiv \VEV{\delta_g\delta_g}_{k,z} = b_g(z)^2\Plin(k,z)$]. Finally, $b_v(z)$ is used to account for a bias in the velocity reconstruction. The radial component of this field can be obtained on large scales, using the linear-theory relation $v_r(\bk, z) = ik_rv(\bk,z)/k$. This measurement, therefore, offers an independent tracer of the $\eta$-field [introduced in Eq.~\eqref{eq: K_nhat_eta_model}] that can be cross-correlated with the kSZ-squared data set to supplement the signal from He reionization at low redshifts.

The methodology of velocity reconstruction, as shown above,
was most recently employed in Refs.~\citep{Guachalla:2023lbx,Hadzhiyska:2023nig}, which highlighted several challenges associated with the technique.
First, nonlinear gravitational evolution and the virial motions of satellite galaxies introduce large random velocities (Fingers-of-God) on small scales, which the linear continuity equation cannot fully capture. 
Second, photometric redshift (photo-$z$) errors or residual redshift-space distortions (RSDs) degrade the line-of-sight component of reconstructed velocities. 
Third, uncertainties in the galaxy-halo connection can bias the measurement, as galaxy velocities (especially of satellites) may differ substantially from those of their host halos. 
Fourth, incomplete sky coverage and survey boundaries, as well as a mismatch between the assumed and true cosmology, can introduce further bias and reduce the correlation with true velocities. 
Finally, the need to smooth the density field (to suppress noise and nonlinearities) inevitably discards some velocity information at smaller scales. 
These effects combine to limit or bias the accuracy of velocity reconstruction methods.
We leave a detailed analysis of how this degradation in velocity reconstruction affects the measurement of reionization parameters to future work.

\subsubsection{Radial velocity field squared reconstruction}

Since galaxy survey measurements offer a three-dimensional tracer of the underlying radial velocity field, this data-set can be used to obtain a tomographic measurement of the $\eta(\bn, z)$ field that sources the $L$-dependence of the $C_L^{KK}$ signal from redshift $z$ [see, for example, Eq.~\eqref{eq: C_L_KK_model_expression}]. 
Given the real-space definition of the field $\eta(\bn,z)$ under Eq.~\eqref{eq: K_nhat_eta_model}, it can be obtained from the cosmological velocity field $v(\bk,z)$ as follows:
\begin{eqnarray}
    \eta(\bk) = \frac{1}{\VEV{v_r^2}}\int \frac{d^3\bk'}{(2\pi)^3}\frac{ik_r'}{k'}\frac{i(k_r-k_r')}{|\bk-\bk'|} v(\bk')v(\bk-\bk')\,,
    \label{eq: true_eta_field_def}
\end{eqnarray}
where the $z$ dependence of the fields has been suppressed for ease of notation.

In reality, due to noise in the reconstruction of $v(\bk,z)$, the $\hat{\eta}(\bk,z)$ field \textit{estimated} from observed galaxy-density measurements differs from the \textit{true} underlying $\eta(\bk,z)$ field. 
We define the minimum variance estimator for the galaxy-reconstructed $\hat{\eta}(\bk,z)$ field as follows:
\begin{equation}
    \hat{\eta}(\bk) = \int \frac{d^3\bk'}{(2\pi)^3}\frac{W(k',|\bk-\bk'|)}{\VEV{v_r^2}}\frac{k_r'}{k'}\frac{(k_r-k_r')}{|\bk-\bk'|} \hatv(\bk')\hatv(\bk-\bk')\,,
    \label{eq: MVE_eta}
\end{equation}
where the weights $W(k_1, k_2)$ are chosen to minimize the noise in the reconstruction of the $\eta$-field. 
In other words, the weights $W(k_1, k_2)$ are chosen to minimize $N_{\eta\eta}^{\perp}(k) \equiv \lr{[}{P_{\hateta\hateta}^{\perp}(k) - P_{\eta\eta}^{\perp}(k)}{]}$, 
where $P_{\hateta\hateta}(k,z) \equiv \VEV{\hateta\hateta}_k$ is the \textit{observed} power spectrum of the reconstructed field [also denoted $\td{P}_{\eta\eta}(k,z)$], 
and $P_{\eta\eta}(k)$ is the true power spectrum of the underlying $\eta(\bk,z)$ [Eq.~\eqref{eq: p_eta_perp}]. 
Recall that the the superscript `$\perp$' simply imposes the reconstruction of modes with $k_r = 0$. 

\begin{figure}
    \centering
    \includegraphics[width=\linewidth]{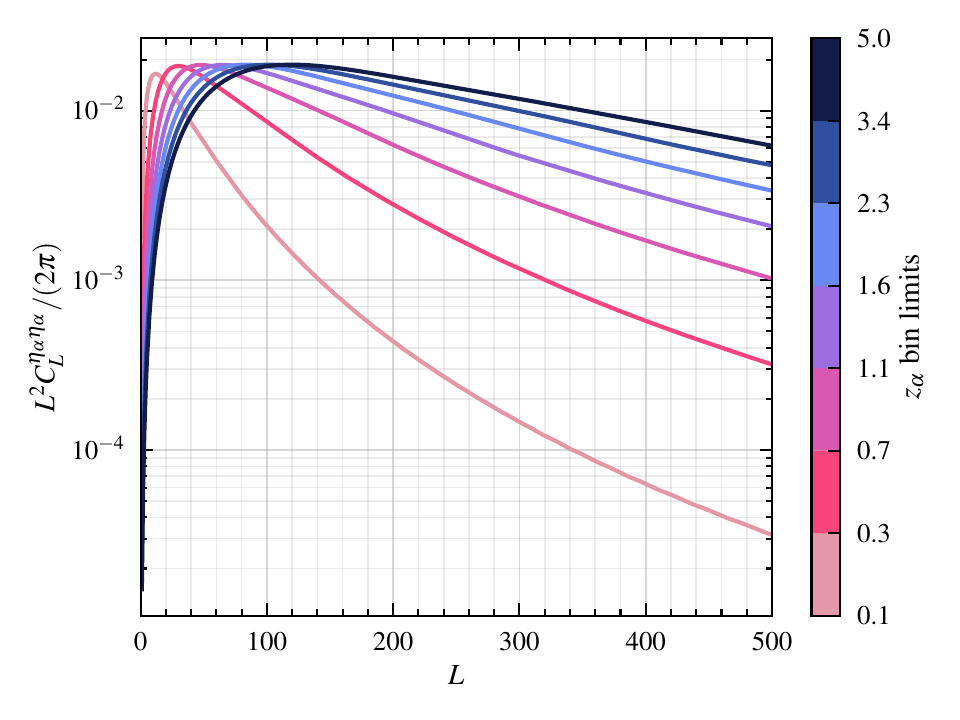}
    \caption{The angular power-spectra $C_L^{\eta\eta}$
    computed for $\Nzbins = 7$ equal-width (in comoving distance) redshift bins between $0.1 \leq z \leq 5.0$. The signal is computed using Eq.~\eqref{eq: C_L_eta_eta_obs_binned}, assuming that the velocity reconstruction bias $b_v$ is 1.0 at all redshifts. The tick markers on the color-bar correspond (approximately) to the bin-limits in redshift-space.
    }
    \label{fig:nn_signal_only_binned}
\end{figure}

Solving for $W(k_1,k_2)$ involves not only minimizing the reconstruction noise $N_{\eta\eta}^{\perp}(k)$, but also ensuring that the re-constructed $\hateta(\bk)$ must simplify to  $\eta(\bk) + n_\eta(\bk)$ where $n_\eta(\bk)$ is the (uncorrelated) noise in the reconstruction [i.e., $N_{\eta\eta}(k) = \VEV{n_\eta n_\eta}_k$].
This constraint is equivalent to requiring $P_{\hateta\,\eta}(k) = P_{\eta\eta}(k)$, where 
\begin{equation}
\begin{aligned}
    P_{\hat\eta\,\eta}(k) = \frac{-2}{{\VEV{v_r^2}}^2} &\int \frac{d^3\bk'}{(2\pi)^3}
    \, \left( \frac{k_r'}{k'}\frac{(k_r - k_r')}{|\bk - \bk'|} \right)^2 \\
    &\quad \times W(k',|\bk - \bk'|)
    P_{vv}(k') P_{vv}(|\bk - \bk'|)\,.
\end{aligned}
\end{equation}
Using the method of Lagrange multipliers, with the goal of minimizing $N_{\eta\eta}^{\perp}(k)$ while subject to the above constraint equation, one can derive:
\begin{equation}
    W(k_1,k_2) = \frac{-P_{vv}(k_1)P_{vv}(k_2)}{\tilde{P}_{vv}(k_1)\td{P}_{vv}(k_2)}\lr{[}{\frac{P_{\eta\eta}(k)}{Q(k)}}{]}_{\bk=\bk_1+\bk_2}\,.
    \label{eq: eta_recon_weights}
\end{equation}
In the above equation, $\td{P}_{vv}(k, z)$ represents the \textit{observed} velocity power spectrum $P_{vv}(k,z) + N_{vv}(k,z)$, where $N_{vv}(k,z)$ is the noise is the galaxy-reconstructed cosmological velocity field. Moreover, in fixing the normalization of $W(k_1, k_2)$ to adhere to the constraint on the estimator $\hateta(\bk,z)$, we have defined
\begin{equation}
\begin{aligned}
    Q(\bk) \equiv \frac{2}{\VEV{v_r^2}^2}\int \frac{d^3\bk'}{(2\pi)^3}
    \, &\left(\frac{k_r'}{k'}\frac{(k_r - k_r')}{|\bk - \bk'|} \right)^2\\ &\times \frac{[P_{vv}(k')P_{vv}(|\bk-\bk'|)]^2}{\tilde{P}_{vv}(k')\td{P}_{vv}(|\bk-\bk'|)}\,.
\end{aligned}
\end{equation}
Assuming that the most dominant source of noise in observations of $P_{gg}(k,z)$ is shot-noise, the noise in the reconstructed cosmological velocity field can be written as
\begin{eqnarray}
    N_{vv}(k,z) \equiv \left[\frac{f(z)H(z)}{k(1+z)}\frac{b_v(z)}{b_g(z)}\right]^2\frac{1}{n_g(z)}\,,
\end{eqnarray}
where $n_g(z)$ is the mean galaxy number density at redshift $z$. 

Finally, plugging the derived $W(k_1, k_2)$ [Eq.~\eqref{eq: eta_recon_weights}] into the initial definition of the minimum-variance estimator in Eq.~\eqref{eq: MVE_eta} allows us to derive an expression for the reconstruction noise in the galaxy-estimated $\hat{\eta}(\bk,z)$:
\begin{widetext}
\begin{eqnarray}
    N_{\eta\eta}(k) = \frac{2}{\VEV{v_r^2}^2}\frac{P_{\eta\eta}(k)}{Q(k)}\int\frac{d^3\bk'}{(2\pi)^3}\left(\frac{k_r'}{k'}\frac{(k_r - k_r')}{|\bk - \bk'|} \right)^2
    \frac{P_{vv}(k')P_{vv}(|\bk-\bk'|)}{\td{P}_{vv}(k')\td{P}_{vv}(|\bk-\bk'|)}[\td{P}_{vv}(k')\td{P}_{vv}(|\bk - \bk'|) - P_{vv}(k')P_{vv}(|\bk-\bk'|)]\,,\nonumber \\
    \label{eq: N_eta_perp}
\end{eqnarray}
\end{widetext}
where the $z$ dependence of all the power spectra has been suppressed for ease of notation.

The three-dimensional reconstructed field $\hateta(\bn,z)$ can then be converted to a set of binned angular power spectra as follows. The (redshift-bin averaged) radial-velocity-power in a bin centered at $z_\alpha$ is given by:
\begin{eqnarray}
    \hateta_\alpha(\bn) = \frac{1}{\Delta\chi_\alpha}\int_{z_\alpha^{\rm min}}^{z_\alpha^{\rm max}} \frac{\dd z}{H(z)}\frac{\hatv_r(\bn, z)^2}{\VEV{v_r(z)^2}}\,,
    \label{eq: eta_alpha_gal_recon}
\end{eqnarray}
where  $\Delta\chi_\alpha$ is the comoving width of the bin, and $[z_\alpha^{\rm min}, z_\alpha^{\rm max}]$ are its limits.
This method, therefore, allows for the construction of a set of fields $\{\eta_\alpha\}$ for $\alpha \in [1,\,\Nzbins]$, where $\Nzbins$ is the number of redshift bins the galaxy survey data is divided into.
The \textit{observed} power spectrum $\tilde{C}_L^{\eta_\alpha\eta_\beta}$, sourced across two redshift bins centered at $z_\alpha$ and $z_\beta$, can then be written as:
\begin{eqnarray}
    \tilde{C}_L^{\eta_\alpha\eta_\beta} &= &C_L^{\eta_\alpha\eta_\beta} + \delta_{\alpha\beta}N_L^{\eta_\alpha\eta_\alpha}\nonumber\\
    &=&\frac{\delta_{\alpha,\beta}}{\Delta \chi_\alpha^2}\int_{z_{\rm min}^\alpha}^{z_{\rm max}^{\alpha}} \frac{\dd z}{H(z) \chi(z)^2}\tilde{P}_{\eta\eta}^{\perp}(L/\chi, z)\,,\nonumber\\
    \label{eq: C_L_eta_eta_obs_binned}
\end{eqnarray}
where $N_L^{\eta_\alpha\eta_\alpha}$ is the noise in the reconstructed angular power spectrum (expected to be zero for $\alpha \neq \beta$), and in the second line we have used the Limber approximation and assumed minimal correlation between $\eta_\alpha(\bn)$ and $\eta_\beta(\bn)$ for $\alpha\neq\beta$. 
Finally, $\tilde{P}_{\eta\eta}^{\perp}(k,z) \equiv P_{\hateta\hateta}(k,z) = P_{\eta\eta}^\perp(k,z) + N_{\eta\eta}^{\perp}(k,z)$, where the three-dimensional signal and noise power spectra are provided in Eqs.~\eqref{eq: p_eta_perp} and~\eqref{eq: N_eta_perp} (evaluated for $k_r = 0$). 

The dependence of the signal $C_L^{\eta_\alpha\eta_\beta}$ and noise $N_L^{\eta_\alpha\eta_\alpha}$ on galaxy survey parameters $\{b_v(z),\, b_g(z),\, n_g(z)\}$ is revealed by replacing every instance of $P_{vv}(k,z)$ and $N_{vv}(k,z)$ in Eqs.~\eqref{eq: p_eta_perp} and \eqref{eq: N_eta_perp} with $(b_vfH/[b_g(1+z)k])^2 P_{gg}(k,z)$ and $(b_vfH/[b_g(1+z)k])^2 n_g^{-1}$, respectively. For the forecasts that follow, we assume that our galaxy survey/s of choice allow for $\Nzbins = 7$ redshift bins across $0.1 \leq z \leq 5.0$. Figure~\ref{fig:nn_signal_only_binned} displays the signal $C_L^{\eta_\alpha\eta_\alpha}$ calculated using Eq.~\eqref{eq: C_L_eta_eta_obs_binned}, assuming that the velocity reconstruction bias $b_v = 1.0$ at all redshifts. The tick labels of the color-bar approximately define the limits of each equal-width bin considered.

\subsubsection{Cross-correlation angular power spectrum}
\begin{figure*}
    \centering
    \includegraphics[width=\linewidth]{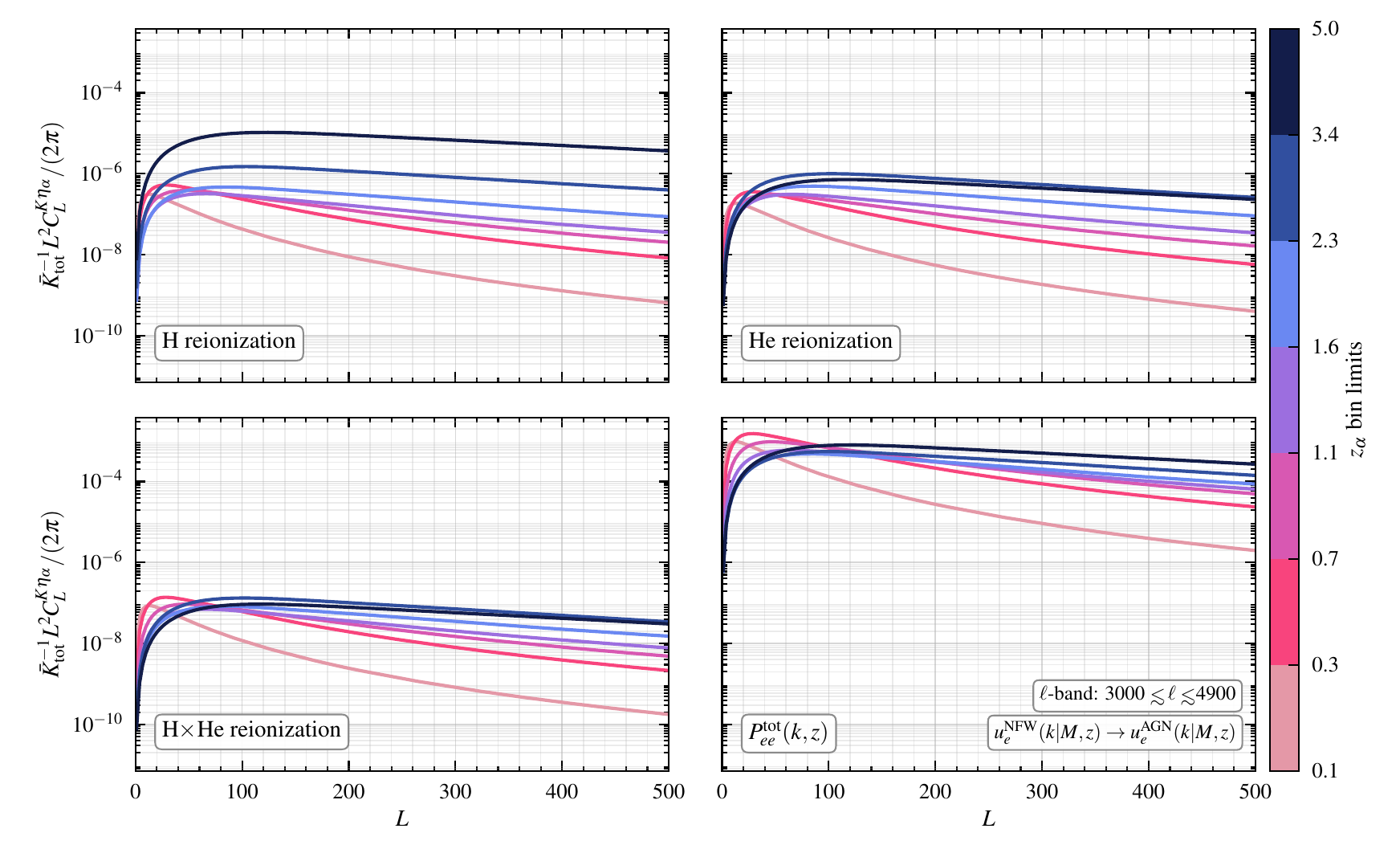}
    \caption{Separate contributions to the binned $C_L^{K\eta_\alpha}$ angular power spectrum for $\alpha \in [1,\, \Nzbins = 7]$. 
    The results have been computed in the same equal-width redshift bins used in Fig.~\ref{fig:nn_signal_only_binned}, with $K(\bn)$ obtained from $3000 \lesssim \ell \lesssim 4900$. The top-left (top-right) panel displays the contribution from patchy H reionization (He reionization). This includes contributions from terms in Eq.~\eqref{eq: ion_elec_PS_as_ft_2pt_terms} that may additionally depend on $\netot(\br,z)$ but do not depend on both $\xH(\br,z)$ and $\xHe(\br,z)$. The contribution from these `mixed' terms is presented in the bottom-left panel of the figure, labeled `H$\times$He reionization'. Finally the contribution sourced solely by $P_{ee}^{\rm tot}(k,z)$ is presented in the bottom-right panel. All the signals are computed using an AGN profile for $u_e(k|M,z)$, and assuming that $C_\ell^{\rm tot} = C_\ell^{TT} + C_\ell^{\rm kSZ} + N_\ell^{\rm HD} + F_\ell^{\rm HD}$ (see Sec.~\ref{subsec: Experiment Specifications} for further details on $N_\ell^{\rm HD}$ and $F_\ell^{\rm HD}$). Note that the signals have been normalized by $\bar{K}_{\rm tot}$ for ease of comparison to the $C_L^{KK}$ signal presented in Fig.~\ref{fig:KK_signal_and_noise_all_parts_with_dz}. 
    }
    \label{fig:Kn_signal_only_all_parts}
\end{figure*}

Given that galaxy surveys can be used to reconstruct the set of fields $\{\eta_\alpha(\bn)\}$, this additional tracer can then be cross-correlated with with each of the $\ell$-binned kSZ-squared fields $\{K_i(\bn)\}$ to obtain the $\Nellbins \times \Nzbins$ cross-correlation signals:
\begin{eqnarray}
    C_L^{K_i\eta_\alpha}=\frac{1}{\Delta \chi_\alpha}\int_{z_{\rm min}^\alpha}^{z_{\rm max}^{\alpha}} \frac{\dd z}{\chi(z)^2} \frac{\dd \bar{K}_i(z)}{\dd z}\,P_{\eta\eta}^\perp(k=L/\chi)\,.
    \label{eq: C_L_K_eta}
\end{eqnarray}
In the above equation, we have once again assumed that the correlation between the galaxy-reconstructed  $\eta_\alpha(\bn)$ and velocity field sourcing the kSZ effect outside the limits $[z_{\rm min}^\alpha, z_{\rm max}^\alpha]$ is minimal. 
From this point on, whenever describing the general characteristics of the $\Nellbins \times \Nzbins$ cross-correlation signals, we will refer to the set with shorthand $C_L^{K\eta}$ or simply $K\eta$.

Figure~\ref{fig:Kn_signal_only_all_parts} displays the separate contributions to the total $C_L^{K_1\eta_\alpha}$ signal, computed for $3000 \lesssim \ell \lesssim 4900$ in the same seven equal-width redshift bins used in Fig.~\ref{fig:nn_signal_only_binned}. 
Consistent with Fig.~\ref{fig:KK_signal_and_noise_all_parts_with_dz}, in each redshift bin centered at $z_\alpha$, the total $C_L^{K_1\eta_\alpha}$ is separated into four contributors. 
The top-left (top-right) panel displays the contribution from terms in Eq.~\eqref{eq: ion_elec_PS_as_ft_2pt_terms} [propagated through to Eq.~\eqref{eq: C_L_K_eta}] that are dependent on characteristics of H reionization (He reionization). 
These terms may include a dependence on $\netot(\br,z)$, but are not dependent on \textit{both} $\xH(\br,z)$ and $\xHe(\br,z)$. 
The contribution from these `mixed' terms, that are sourced by fluctuations in both $\xH(\br,z)$ and $\xHe(\br,z)$, is presented in the bottom-left panel. 
Finally, the bin-wise contribution from the first term in Eq.~\eqref{eq: ion_elec_PS_as_ft_2pt_terms}, solely dependent on fluctuations in $\netot(\br,z)$, is presented in the bottom-right panel of the figure. 
To remain consistent with the auto-correlation $C_L^{K_1K_1}$ plotted in Fig.~\ref{fig:KK_signal_and_noise_all_parts_with_dz}, the $C_L^{K_1\eta_\alpha}$ curves are computed assuming an AGN profile for $u_e(k|M,z)$ and are normalized by $\bar{K}_{\mathrm{tot}, 1}$ for ease of comparison. 
Furthermore, $\dd \bar{K}_1(z)/ \dd z$ in Eq.~\eqref{eq: C_L_K_eta} is computed using Eq.~\eqref{eq: K_nhat_eta_model} assuming that $C_\ell^{\rm tot} = C_\ell^{TT} + C_\ell^{\rm kSZ} + N_\ell^{\rm HD} + F_\ell^{\rm HD}$. 
Further details on CMB instrument noise $N_\ell^{Y}$ and foreground $F_\ell^Y$ modeling, for both $Y\in \{$S4, HD$\}$, can be found in  Sec~\ref{subsec: Experiment Specifications}.

A noteworthy feature of the cross-correlation signals is that at lower redshifts the contribution to $C_L^{K\eta}$ from He reionization is comparable to the contribution from H reionization. 
This stands in contrast to the large difference in these separate contributors to the auto-power spectrum $C_L^{K_1K_1}$ plotted in Fig.~\ref{fig:KK_signal_and_noise_all_parts_with_dz}, confirming that the cross-correlation probe will be specifically helpful in characterizing patchy He reionization.

Moreover, the  redshift evolution of the curves plotted in each of the four panels of Fig.~\ref{fig:Kn_signal_only_all_parts} is visibly different. 
The distinct evolutions of the separate contributors to $C_L^{K\eta}$ can be explained by carefully considering the redshift evolution of both the epochs of reionization. 
As anticipated from the evolution of $\bxH(z)$ (Fig.~\ref{fig:reionization_simplified_models}), Fig.~\ref{fig:Kn_signal_only_all_parts} confirms that the contribution from terms that are solely dependent on H reionization steadily decreases with decreasing $z_\alpha$, as H in the IGM becomes more uniformly ionized .
Similarly, as the power in $P_{ee}^{\rm tot}(k,z)$ increases with decreasing $z_\alpha$, its contribution to the cross-correlation signal steadily increases. 
In contrast, the behavior of terms dependent on He reionization parameters is less trivial. Curves in both the top-right and bottom-left panel of the figure indicate that the contribution from terms dependent on $\xHe(\br,z)$ is highest for bins in the range $1.1 \lesssim z \lesssim 3.4$. 
This is consistent with the model for $\bxHe(z)$ (Fig.~\ref{fig:reionization_simplified_models}), which predicts maximal patchiness from bubbles of ionized He in this range of redshifts. 

The unique redshift evolution of each of the separate contributors to $P_{ee}^{\rm ion}(k,z)$ is an important characteristic that, if leveraged, can significantly improve our chances of disentangling the effects of patchy reionization from the clustering of $\netot(\br,z)$. 
Furthermore, it can also alleviate any potential degeneracies between the two separate epochs of reionization that are otherwise governed by very similar physical processes. 
Since a tomographic measurement of this evolution is harder to access using the line-of-sight integrated probe $\{K_i(\bn)\}$, the fields $\{\eta_\alpha(\bn)\}$ obtained from the three-dimensional galaxy-survey dataset become an essential tool in disentangling the effects of the two separate epochs. 

\section{Forecasts}
\label{sec:Forecasts}
Given the model for patchy reionization incorporated into $P_{ee}^{\rm ion}(k,z)$ in Sec.~\ref{sec: Reionization Model for the Ionized Electron Power Spectrum}, and the expressions for the auto- and cross-correlation signals $\{C_L^{K_iK_j},\, C_L^{\eta_\alpha\eta_\alpha},\, C_L^{K_i\eta_\alpha}\}$ and noise terms $\{N_L^{K_iK_j},\, N_L^{\eta_\alpha\eta_\alpha}\}$ for $i,j \in [1,\, \Nellbins]$ and $\alpha \in [1,\, \Nzbins]$ presented in Sec.~\ref{sec: kSZ Trispectrum and Galaxy Cross-Correlation}, we now proceed to forecast our ability to characterize patchy reionization with upcoming CMB and galaxy surveys.
In Sec.~\ref{subsec: Experiment Specifications}, we  provide an overview of the experiment specifications for each type of CMB experiment and galaxy survey we consider in the final forecasts. We not only detail instrument specifics but also address any additional sources of noise or foregrounds considered in final analysis. In Sec.~\ref{subsec: Detection SNR}, we proceed to present the estimated measurement SNR from each of the established baselines, presenting results on both the $KK$ auto-correlation signal as well as the $K\eta$ cross-correlation signal. Finally, Sec.~\ref{subsec: Forecasting Parameter Errors}, we delve into the measurability of H and He reionization parameters, assessing potential parameter degeneracies and measurement strategies to optimize future constraints.

\subsection{Experiment Specifications}
\label{subsec: Experiment Specifications}

For the forecasts that follow on both the auto-correlation signal $C_L^{KK}$ and the binned-cross-correlation signal $C_L^{K\eta}$, we assume that the CMB temperature maps are obtained from upcoming surveys---CMB-S4~\cite{CMB-S4:2016ple, Abazajian:2019eic} and CMB-HD~\cite{Sehgal:2019ewc, CMB-HD:2022bsz}. 
For each of these experiments, we model the instrument noise in the observed CMB signal as follows:
\begin{eqnarray}
    N_\ell =\Delta_T^2\exp\left[\frac{\ell(\ell+1)\theta^2_{\rm FWHM}}{8\ln2}\right]\,,
    \label{eq: CMB_red_white_noise}
\end{eqnarray}
where $\Delta_T$ is the detector RMS noise, and $\theta_{\rm FWHM}$ is the Gaussian beam full width at half maximum. 
These parameters are dependent on the frequency at which the CMB temperature map is obtained. 
Their assumed values for all following forecasts, for both the baselines considered, are summarized in Tab.~\ref{tab:beamnoise}. 
We label the white-noise spectra for CMB-S4 and CMB-HD, obtained using the frequency maps listed in Tab.~\ref{tab:beamnoise} post ILC-cleaning, $N_\ell^{\rm S4}$ and $N_\ell^{\rm HD}$ respectively.
Finally, we assume that the smallest scale accessible by CMB-S4 (CMB-HD) is $\ell_{\rm max} = 12000$ ($\ell_{\rm max} = 22000$), with $\Nellbins = 7$ ($\Nellbins = 10$). Although this amounts to fewer $\ell$-bins than considered in Ref.~\cite{Ferraro:2018izc}, we seek to simply demonstrate the specific advantages of binning in multipoles when probing two epochs of reionization simultaneously. That is, the values of $\Nellbins$ are not optimized for parameter estimation.

\begin{table}
\caption{{\it Inputs to ILC noise for the baseline CMB configurations:} 
The parameters $\{\Delta_T,\,\theta_{\rm FWHM}\}$ at each frequency, characterizing instrument white-noise, are chosen to match CMB-S4~\citep{CMB-S4:2016ple} and CMB-HD~\citep{CMB-HD:2022bsz}. The noise spectrum at a given frequency is modeled using Eq.~\eqref{eq: CMB_red_white_noise}. }
\label{tab:beamnoise}
\begin{center}
\renewcommand{\arraystretch}{1.4} 
\setlength{\tabcolsep}{12pt}     
\centering
\begin{tabular}{lcccc}
\hline \hline
\ \  & \multicolumn{2}{c}{Beam FWHM} \ \ & \multicolumn{2}{c}{Noise RMS} \\
& \multicolumn{2}{c}{$\theta_{\rm FWHM}$} & \multicolumn{2}{c}{ $\Delta_T$ ($\mu$K-arcmin)} \\ 
\hline
& S4 & HD & S4 & HD \\ \hline
39 GHz  \   \             & $5.1'$                  & $36.3''$       \    \                          & 12.4                     & 3.4                     \\
93 GHz \   \              & $2.2'$                  & $15.3''$        \     \                         & 2.0                     & 0.6                     \\
145 GHz   \   \           & $1.4'$                  & $10.0''$       \     \                         & 2.0                     & 0.6                     \\
225 GHz \   \            & $1.0'$                  & $6.6''$       \      \                          & 6.9                     & 1.9                     \\
280 GHz  \  \             & $0.9'$                  & $5.4''$      \     \                           & 16.7                    & 4.6                     \\ \hline \hline
\end{tabular}
\end{center}
\end{table}

In addition to the inclusion of instrument noise, we also account for residual contamination from foregrounds in the observed CMB signal. 
We include the frequency-dependent thermal Sunyaev-Zel'dovich (tSZ) effect, computed according to Refs.~\cite{Madhavacheril:2017onh, Park:2013mv}. 
We also consider Poisson and clustered cosmic infrared background (CIB), however we do not account for the tSZ$\times$CIB cross-correlation. 
We also include radio-point sources, following Ref.~\cite{Lagache19}. 
We assume that galactic foreground contamination is removed from the CMB temperature maps, and to remain consistent with this assumption we assume a sky-coverage fraction of $f_{\rm sky} = 0.4$. 
Finally, the lensed primary CMB signal $C_\ell^{TT}$ is calculated using \texttt{CAMB}~\citep{CAMB}. 
In the forecasts that follow, we present results for the full ILC-cleaned CMB signal, assuming that $C_\ell^{\rm tot} = C_\ell^{TT} + C_\ell^{\rm kSZ} + N_\ell^{Y} + F_\ell^{Y}$ for $Y\in$\{S4, HD\} where $F_\ell^{Y}$ represents the residual contamination from all frequency-dependent foregrounds post-ILC cleaning.  

Although the above described foregrounds are incorporated into the observed CMB power spectrum, we neglect the non-Gaussian effects of some foregrounds on the observed $\td{C}_L^{KK}$ signal. One such contribution is sourced by power along the line of sight whose amplitude is linear in the matter over-density field.
Reference~\cite{Smith:2016lnt} establishes that the contribution from this effect, sourced during reionization and at late-times, as well as by residual CIB, is suppressed relative to the kSZ contribution. 
Residual tSZ from un-clustered Poisson sources induces some non-Gaussian signal in the observed $C_L^{KK}$, roughly modeled as a constant offset for large $L$~\cite{Smith:2016lnt, Ferraro:2018izc}. 
Given the high dimensionality of our parameter space, which includes bias parameters $\{b_v,\, b_g,\, b^{\rm H},\, b^{\rm He}\}$, we do not account for nuisance parameters to marginalize over this shot-noise effect.
Furthermore, CMB-lensing  can impart a trispectrum signal comparable to the reconstruction noise $N_L^{KK}$. Although we do not account for this effect in our forecasts, Ref.~\cite{MacCrann:2024ahs} shows that bias hardening can effectively remove this bias in the kSZ-sourced trispectrum signal measurement. Finally, we also neglect the non-Gaussian cross-correlation between the CIB and tSZ foregrounds which may be important to account for in order to make an unambiguous detection in the future~\citep[see e.g.,][]{MacCrann:2024ahs,SPT-3G:2024lko}.

For the forecasts that involve the galaxy-reconstructed, binned $\eta_\alpha(\bn)$ field, we assume the galaxy catalogs are obtained from surveys with specifications matching the upcoming LSST~\citep{2009arXiv0912.0201L} and MegaMapper~\cite{Schlegel:2019eqc} surveys. Since the ``MegaMapper'' concept has now evolved into the broader Stage-5 spectroscopic survey (Spec-S5) framework, one can take MegaMapper to represent a plausible Spec-S5-like survey~\cite{Spec-S5:2025uom}.

We approximate the (average) galaxy number-density in each redshift bin for the anticipated LSST ``gold sample'' using the following equation:
\begin{eqnarray}
    n_\text{gal}(z_\alpha) = n_0\left[\frac{z_\alpha}{z_0}\right]^2\frac{\exp(-z_\alpha/z_0)}{{2z_0}}\,,
\end{eqnarray}
where $n_0=40~\text{arcmin}^{-2}$, and $z_0=0.3$. The above equation provides the number-density of galaxies per-${\rm arcmin}^2$, which is then converted to the required volume-density of galaxies $n_g(z_\alpha)$, assuming that $n_\text{gal}$ takes a fixed value in a given redshift bin centered at $z_\alpha$. 
Moreover, we assume that the galaxy bias takes a fixed value in each bin, which can be approximated using $b_g(z_\alpha)=0.95(1+z_\alpha)$. 
Given that the MegaMapper survey is anticipated to be a high-redshift follow-up on the LSST catalog, we use the above described formalism to estimate $n_{\rm g}(z_\alpha)$ and $b_g(z_\alpha)$ at low redshifts ($z\lesssim 2.0$). For higher-redshift bins, we interpolate the expected galaxy number-density and linear galaxy bias from Tab.~I of Ref.~\cite{Hotinli:2022jna}.
As depicted in Figs.~\ref{fig:nn_signal_only_binned} and ~\ref{fig:Kn_signal_only_all_parts}, we use $\Nzbins = 7$, equal width (in comoving distance) redshift bins within the range $z \in [0.1, 5.0]$.

It is important to note that, in the forecasts presented below, we do not explicitly model the effects of photo-$z$ errors.
The LSST ``gold sample'', considered in this work, is used for most cosmological analyses and has good photometric redshifts associated with each galaxy, which is essential for the work proposed here. 
This sample is primarily comprised of galaxies at $z < 2$. 
Although LSST will also detect higher redshift galaxies (using, for example, dropout techniques), these may have too poor of a photo-$z$ determination to be usable here.
Analyses of synthetic galaxy catalogs on the light-cone, presented in Refs.~\cite{Hadzhiyska:2023nig, Guachalla:2023lbx}, indicate a factor of $\sim 2$ reduction in SNR of velocity reconstruction from LSST-like photo-$z$ errors at low redshift.
To minimize the sensitivity of our forecasts to these errors, we use galaxy bin sizes that are much larger than the anticipated width $\sigma_z=0.03(1+z)$ of photometric redshift errors for a survey like LSST.
In contrast, a spectroscopic follow-up to LSST, such as MegaMapper, aims to get spectroscopic redshifts for large numbers of $z > 2$ galaxies, making this data-set more suitable for our analysis. To account for this, we assume that MegaMapper will have a higher effective $n_{\rm gal}$ at high redshifts.
In summary, therefore, the assumptions made here are optimistic for LSST and the relative improvement from MegaMapper conservative.

\subsection{Detection SNR}
\label{subsec: Detection SNR}
To estimate the sensitivity of upcoming CMB surveys to large-scale clustering in the locally measured kSZ power, we start by estimating the SNR of the binned $KK$ auto-correlation signal detailed in Sec.~\ref{subsec: The kSZ Trispectrum} using the following equation:
\begin{align}
    &\text{SNR}^2 = \sum_{L = 1}^{L_{\rm max}} \sum_{ijmn}^{N_{\ell\text{-bins}}}
    C_L^{K_iK_j} \notag \\
    &\hspace{2.5em} \times\, \textbf{cov}^{-1}\left( \tilde{C}_L^{K_iK_j},\, \tilde{C}_L^{K_mK_n} \right)
    C_L^{K_mK_n}\,,
    \label{eq:auto_corr_SNR_expr}
\end{align}
where $L_{\rm max}$ is the largest accessible multipole in the $K(\bn)$ field, and the indices $\{i,\, j,\, m,\, n\} \in [1,\, N_{\ell\text{-bins}}]$ label the $\ell$-band used to obtain each $K(\bn)$ field. In the above expression, $\tilde{C}_{\smash{L}}^{K_i\, K_j} \equiv C_{\smash{L}}^{K_i\, K_j} + N_{\smash{L}}^{K_i\, K_j}$, where the expressions for the binned $C_{\smash{L}}^{K\, K}$ and $N_{\smash{L}}^{K\, K}$ are provided in Eqs.~\eqref{eq: C_L_KK_model_expression} and \eqref{eq: N_L_KK_expression}, respectively.

Under similar notational conventions, we also forecast the expected measurement SNR of the $\Nellbins \times \Nzbins$ cross-correlation signals $C_L^{K\eta}$ using both (upcoming) CMB experiments and galaxy-survey data as follows:
\begin{align}
    {\rm SNR}^2 &= \sum_{L=1}^{L_{\rm max}} \sum_{ij}^{\Nellbins} \sum_{\alpha\beta}^{\Nzbins} 
    C_L^{K_i\eta_\alpha} \notag \\
    &\hspace{1.5cm} \times \textbf{cov}^{-1} 
    \left(\tilde{C}_L^{K_i\eta_\alpha},\,\tilde{C}_L^{K_j\eta_\beta}\right) C_L^{K_j\eta_\beta}\,,
    \label{eq: cross_corr_SNR_expr}
\end{align}
where indices $\alpha$ and $\beta$ label redshift bins (each centered at $z_\alpha$ and $z_\beta$, respectively). The terms $\tilde{C}_L^{K_i\eta_\alpha}$ and $\tilde{C}_L^{K_j\eta_\beta}$are computed using Eqs.~\eqref{eq: C_L_eta_eta_obs_binned}-\eqref{eq: N_eta_perp}. The sum over multipoles $L$ spans the same range as in Eq.~\eqref{eq:auto_corr_SNR_expr}. 

In both the quoted SNR expressions, the covariance matrix can be computed using the following equation:
\begin{eqnarray}
    \textbf{cov}\left(\tilde{C}_L^{XY},\,\tilde{C}_L^{WZ}\right) = \frac{\tilde{C}_L^{XW}\tilde{C}_L^{YZ} + \tilde{C}_L^{XZ}\tilde{C}_L^{YW}}{f_{\rm sky}(2L+1)}\,,
\end{eqnarray}
where we have implicitly assumed that the different multipoles $L$ are uncorrelated. 
\begin{figure}
    \centering
    \includegraphics[width=\linewidth]{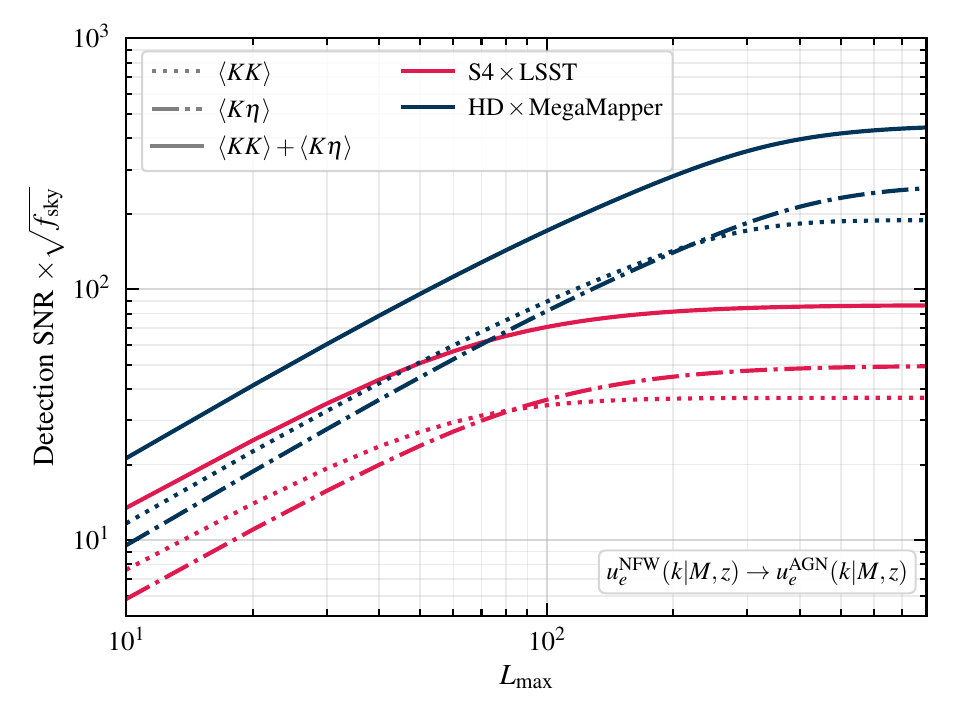}
    \caption{
    Detection SNR as a function of the maximum accessible multipole $L_{\rm max}$ in the fields $K(\bn)$ and $\eta(\bn)$, illustrating the statistical power of upcoming CMB and galaxy surveys in detecting non-Gaussian contributions to the $K(\bn)$ field sourced by the kSZ effect.
    The survey sky-coverage fraction $f_{\rm sky}$ is set to unity.
    The dotted [dot-dashed] lines labeled $\langle KK \rangle$ [$\langle K \eta \rangle$] indicate the detection SNR from the $\ell$-binned $KK$ auto-correlation [$(\ell\times z)$-binned $K\eta$ cross-correlation], while the solid lines combine both signals.
    The pink (blue) line labeled `S4 $\times$ LSST' (`HD $\times$ MegaMapper') assumes a CMB-S4-like (CMB-HD-like) survey cross-correlated with LSST (MegaMapper). Results are computed using the NFW (AGN) profile for $u_e(k|M,z)$ at $z \gtrsim 5$ ($z \lesssim 5$).
    }
    \label{fig:SNR_KK_Kn_ellBin_total_NFW_to_AGN}
\end{figure}

Figure~\ref{fig:SNR_KK_Kn_ellBin_total_NFW_to_AGN} displays the forecasted detection SNRs from the auto- and cross-correlation signals, as a function of $L_{\rm max}$, for both the baselines described in Sec.~\ref{subsec: Experiment Specifications}. Here, $f_{\rm sky}$ is set to unity. 
The estimated $\ell$-binned $KK$ auto-correlation SNR for measurements of the kSZ effect from CMB-S4 and CMB-HD are displayed by the dotted pink and blue curves, respectively. 
In contrast, the dot-dashed curves correspond to the forecasted SNR from the $(\ell\times z)$-binned $K\eta$ cross-correlation  signals, accounting only for redshift bins with $z \lesssim 5$. 
The pink (blue) dot-dashed curve represents the forecast for a CMB-S4-like (CMB-HD-like) experiment cross-correlated with velocity-squared reconstruction from a galaxy-survey matching LSST (MegaMapper). The solid curves represent the total SNR expected from the two independent probes. All the displayed curves have been computed assuming an NFW (AGN) profile for $\netot(\br, z)$ at $z\gtrsim 5$ ($z\lesssim 5$). 

For both the cross-correlation and the auto-correlation, our forecasted detection SNRs increase with the addition of information at higher multipoles (increasing $L_{\rm max}$). 
However, the improvement plateaus at $L_{\rm max} \sim 100$ ($L_{\rm max} \sim 300$) for the CMB-S4/ CMB-S4 $\times$ LSST (CMB-HD/ CMB-HD $\times$ MegaMapper) baseline. 
In the case of the $KK$ auto-correlation forecasts, the plateau can be explained by considering the signal and noise plotted in top panel of Fig.~\ref{fig:KK_signal_and_noise_all_parts_with_dz}. 
Because $C_L^{KK}$ is a large-scale probe, i.e., its $L$-dependence is characterized by $P_{\eta\eta}^\perp(k)$ [Eq.~\eqref{eq: p_eta_perp}] which decreases steeply beyond $k\gtrsim 10^{-1}\,{\rm Mpc}^{-1}$, the $z$-integrated signal drops after $L \gtrsim 200$. 
Moreover, the noise $N_L^{KK}$ in the observed increases following a power law, surpassing the signal at $L \sim 100$. 
The threshold for plateau differs across CMB-S4 and CMB-HD due to the improved noise profile assumed for the latter survey. 
A similar explanation applies to the $K\eta$ cross-correlation SNR as well. 
However, since the `noise' in the cross correlation is computed using a covariance matrix that not only includes $N_L^{KK}$ but also $N_L^{\eta\eta}$, this signal is able to leverage the more moderate increase of $N_L^{\eta\eta}$ as a function of $L$. 
This results in improved marginal returns on the forecasted SNR with increasing $L_{\rm max}$ and a slightly delayed plateau. 
Finally, it is important to note that although the forecasted SNR from the $KK$ auto-correlation signal and the $K\eta$ cross-correlation signal are roughly similar, the $KK$ SNR is sourced by fluctuations in $\neion(\br, z)$ across \textit{all} redshifts, whereas the $K\eta$ SNR is sourced by these fluctuations only at $z\lesssim 5$. 
That is, the information content about the low redshift ionized electron number density is likely considerably higher in the $K\eta$ cross-correlation signal, making it an ideal supplementary probe to aid in the characterization of He reionization. 

\begin{figure*}
    \centering
    \includegraphics[width=\linewidth]{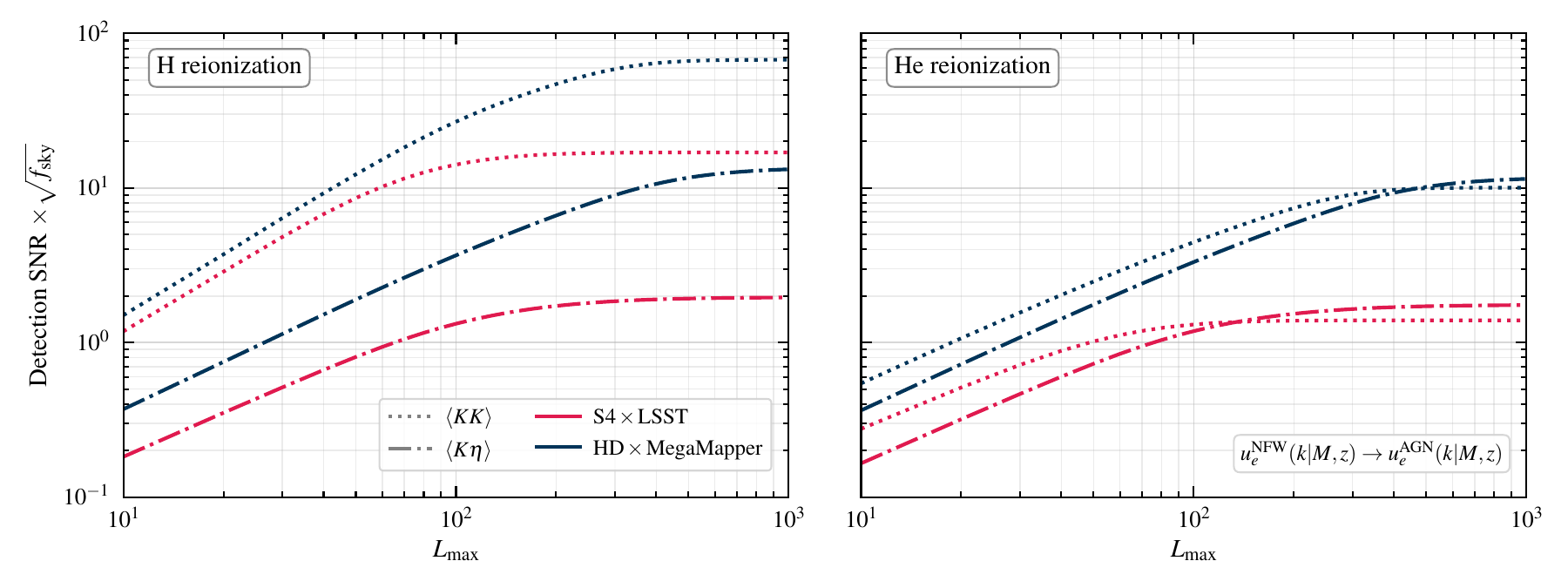}
    \caption{
    Similar to Fig.~\ref{fig:SNR_KK_Kn_ellBin_total_NFW_to_AGN}, but illustrating the detection SNR driven by $x_{\rm H}(\bm{r}, z)$ and $x_{\rm He}(\bm{r}, z)$ in the $\ell$-binned $KK$ and $(\ell \times z)$-binned $K \eta$ signals. The left (right) panel shows contributions from terms dependent on H (He) reionization. The quoted SNRs have been computed assuming $f_{\rm sky} = 1$. Our results indicate that to measure fluctuations with SNR $\gtrsim 3$ in the $K(\bn)$ field sourced by H (He) reionization, assuming $f_{\rm sky} = 0.4$,  one would need to reconstruct the $KK$ auto-correlation signal up to $L_{\rm max} \sim 50$ ($L_{\rm max}\sim200$) for CMB-HD. Similarly, for a detection of H or He reionization with SNR $\gtrsim 3$ in the $K\eta$ cross-correlation signal, one would need to access an $L_{\rm max} \sim 200$ with CMB-HD $\times$ MegaMapper. 
    }
    \label{fig:SNR_KK_Kn_ellBin_H_and_He_NFW_to_AGN}
\end{figure*}

Moreover, to investigate the ability of upcoming surveys to detect fluctuations in $\neion(\br,z)$ sourced \textit{specifically} by the epochs of reionization, Fig.~\ref{fig:SNR_KK_Kn_ellBin_H_and_He_NFW_to_AGN} displays the SNR sourced by $\xH(\br,z)$ and $\xHe(\br,z)$ in the $\ell$-binned $KK$ auto-correlation signal and the $(\ell\times z)$-binned $K\eta$ cross-correlation signal, assuming $f_{\rm sky}=1.0$. 
The left panel specifies the SNR contribution from any terms dependent on H-reionization in Eq.~\eqref{eq: ion_elec_PS_as_ft}. 
Once again, blue and pink dotted curves represent the this contribution to the $C_L^{KK}$-forecasted SNR for CMB-HD and CMB-S4, respectively. 
Moreover, the blue (pink) dot-dashed curves represent the contribution of H-dependent terms to the estimated $C_L^{K\eta}$ SNR, assuming that CMB measurements are obtained from a CMB-HD-like (CMB-S4-like) survey and cross-correlated with MegaMapper (LSST). 
The right panel of the figure represents the same estimations made for the contributions from terms dependent on He-reionization in Eq.~\eqref{eq: ion_elec_PS_as_ft}.

The curves displayed on the left panel of Fig.~\ref{fig:SNR_KK_Kn_ellBin_H_and_He_NFW_to_AGN} show that, as anticipated, the SNR from terms in $P_{ee}^{\rm ion}(k,z)$ sourced by H reionization is larger in the $KK$ auto-correlation signals than in the $K\eta$ cross-correlation signals, for each of the baselines considered. 
Given that fluctuations in $\xH(\br,z)$ are largest at $z\sim 8.5$ (Fig.~\ref{fig:reionization_simplified_models}), this result is a manifestation of the fact that the cross-correlation signals can only probe $\xtot(\br,z)$ up to $z\sim5$. 
In contrast, the estimated SNRs in the right panel of the figure indicate that the sensitivity of the $KK$ auto-correlation signals to the epoch of He reionization is approximately similar to that of the $K\eta$ cross-correlation signals for each baseline. 
This is consistent with the later onset of He reionization, further confirming the fact that the cross-correlation signal will be an essential tool in disentangling the morphologies of the two separate epochs. 

In order to assess the sensitivity of the surveys to He reionization, relative to their ability to probe fluctuations sourced by  H reionization, it is informative to compare the forecasted SNRs across the two panels of Fig.~\ref{fig:SNR_KK_Kn_ellBin_H_and_He_NFW_to_AGN} for each of the signals considered. 
Given a fixed baseline, the forecasted SNRs indicate that the He reionization signal in the $KK$ auto-correlation data-set will be suppressed relative to the H signal by roughly an order of magnitude. 
Superficially, one might expect the SNR from He reionization in auto-correlation signals to be suppressed by two orders of magnitude ($\propto \epsilon^2$). 
However, it is important to note that the He contribution to the $C_L^{KK}$ signals (see, e.g., Fig.~\ref{fig:KK_signal_and_noise_all_parts_with_dz}) includes terms in $P_{ee}^{\rm ion}(k,z)$ that are proportional to $\VEV{\xHe\bnetot}$ that ore only linearly suppressed relative to the $\xH(\br,z)$-dependent terms. 
Moreover, the SNR from He [H] reionization plotted in the right [left] panel of the figure is computed via Eq.~\eqref{eq:auto_corr_SNR_expr}, where \textit{at least} one of the signal vectors $\{C_L^{K_iK_j},\, C_L^{K_mK_n}\}$ is dependent on $\xHe(\br,z)$ [$\xH(\br,z)$]. Therefore, the suppression of the He-dependent terms in forecasted $C_L^{KK}$ SNR is only an order of magnitude relative to the quoted H reionization SNR for each baseline.  
In contrast, the $\xHe(\br,z)$-information content is comparable that of $\xH(\br,z)$ in the $K\eta$ cross-correlation SNRs. This is consistent with the slower model of reionization considered in our work, that still predicts and amplitude of $\bxH(z)$ comparable to He's relative abundance down to $z\sim 4$. 

Finally, because our model for anisotropy in $\neion(\br,z)$ sourced by $\xH(\br,z)$ and $\xHe(\br,z)$ explicitly depends on the small-scale electron power spectrum [1$h$ terms in Eqs.~\eqref{eq: tot_elec_dist_fourier_2pt}-\eqref{eq: ne_tot_x_X_fourier_two_point}], we can also explore the effect that $u_e(k|M,z)$ has on the forecasted detection SNRs. 
To roughly estimate the effect of small-scale power in $P_{ee}^{\rm tot}(k,z)$ at \textit{low} redshifts, we assume $u_e^{\rm NFW}(k,z)$ for $z\gtrsim 5$ and vary $u_e^{Y}(k,z)$ for $Y\in\{$NFW, AGN, $W_e(k)\times$NFW$\}$ at $z\lesssim 5.0$. We find that, in the case of the total and He-dependent forecasts, the SNR becomes suppressed with decreasing power on small scales. 
That is, the AGN and NFW profiles predict the highest SNR, with a visible suppression ($\lesssim$ factor of 2) present in the $W_e(k)\times$NFW scenario. 
In contrast, the $W_e(k)\times$NFW proves to be the most optimistic scenario for the H-dependent SNR ($\sim4$x), likely due to the decrease in $N_L^{KK}$ with no cost to the high-redshift reionization signal. 
In all cases, the change is driven mostly by variation in the auto-correlation SNR. 
We find that changing the high redshift ($z\gtrsim 5$) electron profile has no visible impact on the forecasted SNRs.

\subsection{Forecasting Parameter Errors}
\label{subsec: Forecasting Parameter Errors}
To forecast future prospects of jointly characterizing the two separate epochs of reionization using upcoming CMB-experiment data cross-correlated with galaxy survey measurements, we construct an ensemble-information matrix as follows:
\begin{equation}
    \mathcal{F}_{ij} = \sum_{L = 1}^{L_{\rm max}} \sum_{XYWZ}\frac{\partial C_L^{XY}}{\partial \pi_i} \textbf{cov}^{-1}\lr{(}{\tilde{C}_L^{XY}\tilde{C}_L^{WZ}}{)}\frac{\partial C_L^{WZ}}{\partial \pi_j}.
\end{equation}
We consider two different scenarios in our forecasts---one in which the signals are obtained solely from the CMB maps, i.e., $XY,\, WZ \in \{K_iK_j\}$ for $i,j \in [1,\, \Nellbins]$ and another in which we leverage the cross-correlation with galaxy survey data, i.e., $XY,\, WZ \in \{K_iK_j,\, K_i\eta_\alpha,\, \eta_\alpha\eta_\beta\}$ for $i,j \in [1,\, \Nellbins]$ and  $\alpha,\beta\in[1,\, \Nzbins]$.
Our parameter array includes the ten reionization parameters $y_{\rm re}^X,\, \Delta_y^X,\, \bar{R}^X,\, \sigma_{\ln R}^X,\, \textrm{and } b^X\textrm{ for } X\in\{\textrm{H, He}\}$, as well as a bias term $b_{vg} \equiv b_v/b_g $ that jointly accounts for the velocity-reconstruction bias $b_v$ and the galaxy bias $b_g$. 
The assumed fiducial values of each of the reionization parameters can be found in Tab.~\ref{tab:set_of_params}, and the fiducial value of $b_{vg}$ is computed using the galaxy survey specifications detailed in Sec.~\ref{subsec: Experiment Specifications}, assuming that the velocity reconstruction bias is 1.0 for all $z \leq 5$. Finally, as with the previously presented SNR forecasts, we assume that $u_e(k|M,z)$ follows an NFW (AGN) distribution at $z \gtrsim 5$ ($z\lesssim 5$).

\begin{figure*}
    \centering
    \includegraphics[width=\linewidth]{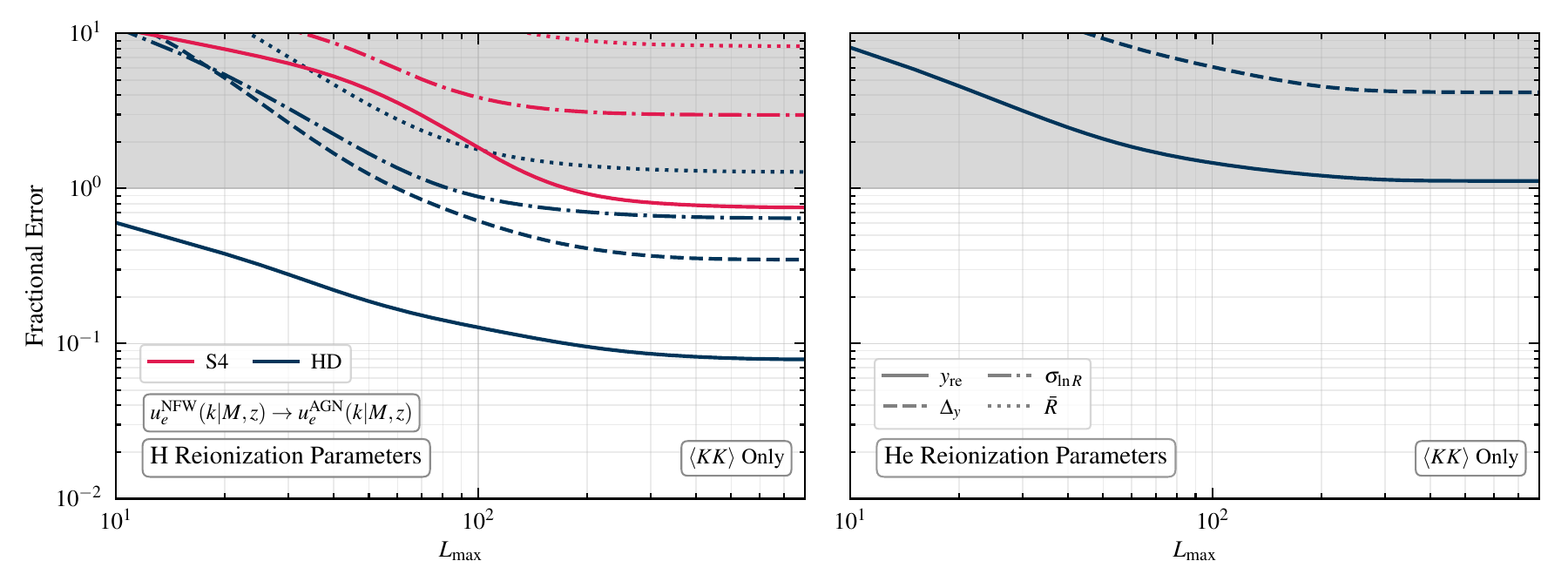}
    \caption{
    Fractional errors on fiducial H and He reionization parameters as a function of maximum multipole $L_{\rm max}$ based only on the $KK$ auto-correlation signal.
    The $y$-axes are fractional errors defined as $\Delta(\pi_i)/\pi_i$, where $\pi \in \{ y_{\rm re}^X,\, \Delta_y^X,\, \bar{R}^X,\, \sigma_{\ln R}^X  \}$, $X \in \{$ H (left panel),  He (right panel)$\}$, and the set of fiducial parameter values are defined in Tab.~\ref{tab:set_of_params}. 
    The $x$-axes correspond to the smallest accessible scale in the field $K(\bn)$.
    The pink (blue) lines represent forecasts obtained assuming experiment specifications corresponding to a CMB-S4-like (CMB-HD-like) survey.
    The gray-shaded region corresponds to fractional errors larger than unity, above which parameters cannot be detected.
    The forecasts indicate that marginalizing over all reionization parameters will only allow for characterization of H reionization, with the mean redshift of reionization $y_{\rm re}^{\rm H}$ detectable within $1\sigma$ by both baselines. Results are computed using the NFW (AGN) profile for $u_e(k|M,z)$ at $z \gtrsim 5$ ($z \lesssim 5$).
    }
    \label{fig:fisher_results_KKonly_ellBin}
\end{figure*}
{\bf \em Auto-correlation forecasts.}
For our first set of forecasts, which rely only on the $KK$ auto-correlation signal, we assume that the set of fields $\{K_i\}$ with $i \in [1,\, N_{\ell\text{-bins}}]$ is constructed by binning the harmonic-space temperature-squared maps within the range $\ell \in [3000,\, \ell_{\rm max}]$ into $N_{\ell\text{-bins}}$ equal-width bands.
We assume $\ell_{\rm max} = 12000$ and $N_{\ell\text{-bins}} = 7$ for CMB-S4, and $\ell_{\rm max} = 22000$ and $N_{\ell\text{-bins}} = 10$ for CMB-HD.
Given the experimental specifications detailed in Sec.~\ref{subsec: Experiment Specifications}, we construct a 10-dimensional information matrix that includes the set of reionization parameters listed previously.
Note that the inclusion of all 10 parameters characterizing the two separate epochs ensures that \textit{our forecasts account for potential parameter degeneracies across the characteristics of H and He reionization}.
Figure~\ref{fig:fisher_results_KKonly_ellBin} summarizes the fractional errors on the fiducial H and He reionization parameters as a function of $L_{\rm max}$.
The pink (blue) curves display the forecasts obtained assuming experiment specifications corresponding to CMB-S4 (CMB-HD).
For each assumed experiment configuration, the solid, dashed, dot-dashed and dotted curves correspond to forecasts on $y_{\rm re}^X,\, \Delta_y^X,\, \bar{R}^X,\, \sigma_{\ln R}^X$, respectively, with the left (right) panel displaying results for $X =$ H ($X$ = He).

The results in Fig.~\ref{fig:fisher_results_KKonly_ellBin} show that the constraining power of each experiment increases with increasing $L_{\rm max}$.
The plateau in marginal improvement from $L_{\rm max} > 200$ is once again explained by the fact that $N_L^{KK}$ steadily increases with $L$ whereas $C_L^{KK}$ has an $L$-dependence is largely attributed to $P_{\eta\eta}^\perp(k)$ [Eq.~\ref{eq: p_eta_perp}], which drops steeply for $k \gtrsim 10^{-1}\,{\rm Mpc}^{-1}$. 
As anticipated, the heavily suppressed contribution from He reionization to the integrated kSZ signal, due to its low relative abundance, leads to poor constraints on any He reionization parameters from both the considered baselines. 
In comparison, even with marginalization over bubble parameters $\bar{R}^{\rm H}$, $\sigma_{\ln R}^{\rm H}$, and $b^{\rm H}$, both baselines are able to constrain $y_{\rm re}^{\rm H}$, which (roughly) sets the mean redshift of H reionization. 

Given that these forecasts are made using the set of $\ell$-binned signals $\{C_L^{K_iK_j}\}$ for $i,j\in[1,\,\Nellbins]$, the sensitivity of a baseline to each H reionization parameter can be explained by considering the parameter's impact of the kSZ effect's $\ell$-dependence. 
For example, both baselines are most sensitive to $y_{\rm re}^{\rm H}$ because a higher value of $y_{\rm re}^{\rm H}$ pushes H reionization to earlier redshifts, increasing the physical electron density contrast between the neutral and ionized regions, thus boosting the power for $3000 \lesssim \ell \lesssim 8000$. 
Furthermore, it also mildly affects the duration of reionization, with larger values of $y_{\rm re}^{\rm H}$ corresponding to shorter epochs of H reionization, leading to a suppression of power for $\ell \gtrsim 9000$. 
Since this mixed effect across scales at $\ell \gtrsim 3000$ is not mimicked by any other parameter, both baselines are able to constrain $y_{\rm re}^{\rm H}$ most effectively. 
Moreover, $\sigma_{\ln R}^{\rm H}$ controls the variance of the ionized bubble-radius distribution and therefore most dramatically impacts the kSZ effect on $\ell \lesssim 7500$.  
As a result, relative to the fractional errors forecasted for $y_{\rm re}^{\rm H}$, each baseline constrains $\sigma_{\ln R}^{\rm H}$ with roughly the same precision. 
In contrast, changing the duration of H reionization causes the CMB photons to encounter more ionized bubbles along the line of sight, and therefore $\Delta_y^{\rm H}$ preferentially boosts power on small scales. 
This effect is more easily leveraged by a CMB-HD-like baseline, that we assume collects data up to $\ell_{\rm max} = 22000$, leading to significantly improved constraints in $\Delta_y^{\rm H}$ relative to a CMB-S4-like survey. 
Finally, although $\bar{R}^{\rm H}$ affects both the shape of $C_{\ell}^{\rm kSZ}$ on $\ell \lesssim 5000$ and the magnitude of power on $\ell \gtrsim 5000$ by changing the average size of ionized regions encountered by the CMB photons, its effect is degenerate with the impact of $\sigma_{\ln R}^{\rm H}$ and $\Delta_y^{\rm H}$ and is relatively subdued. 
As a result, this parameter is measured poorly across both baselines. 
Similarly, the parameter $b^{\rm H}$ is inaccessible by both baselines, since its effect is strongly degenerate with the remaining reionization parameters. 
Finally, it is also important to note that, in the absence of He-reionization and bubble parameters $\bar{R}^{\rm H}$, $\sigma_{\ln R}^{\rm H}$, and $b^{\rm H}$, our forecasts appropriately condense to the forecasts presented in Ref.~\cite{Ferraro:2018izc}. 

\begin{figure*}
    \centering
    \includegraphics[width=\linewidth]{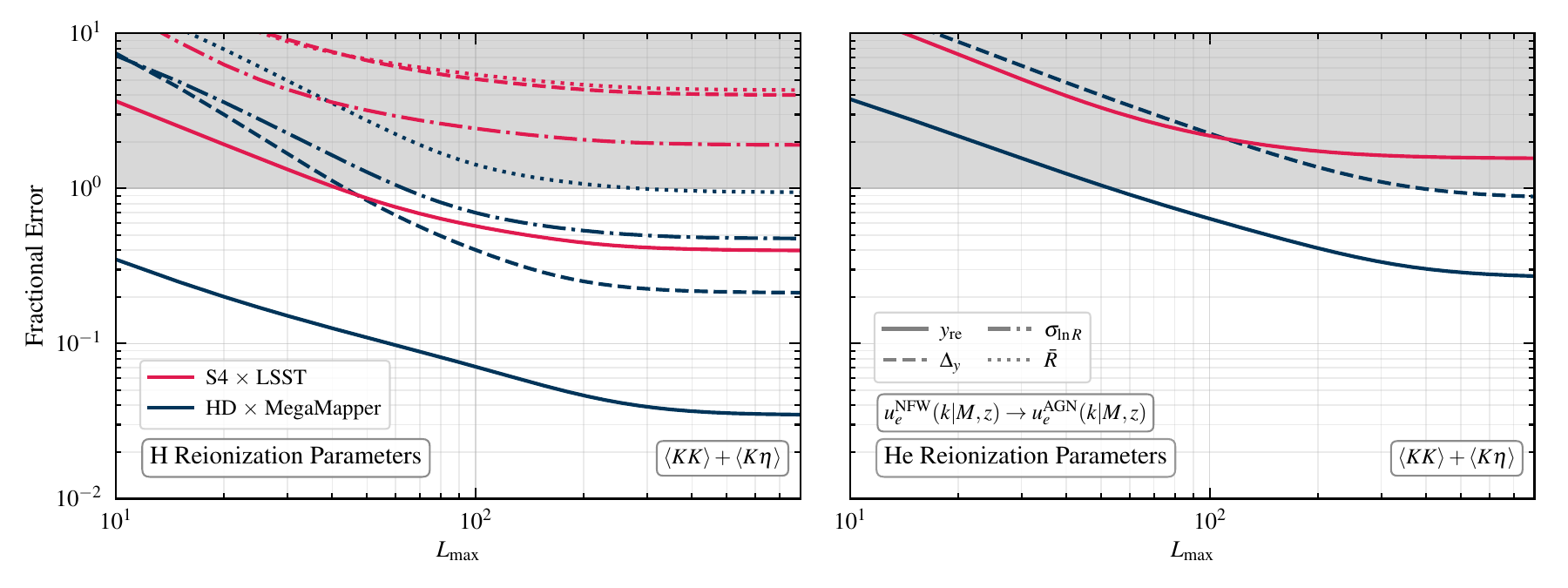}
    \caption{
    Same fractional errors as Fig.~\ref{fig:fisher_results_KKonly_ellBin}, 
    except this set of forecasts accounts for the proposed $K\eta$ cross-correlation between the CMB temperature-squared field and the galaxy-reconstructed radial velocity field, in addition to the $KK$ auto-correlation signal. The information matrix accounts for a galaxy and velocity-reconstruction bias via the parameter $b_{vg}\equiv b_v/b_g$.
    Inclusion of the redshift-binned $\eta$-field data from MegaMapper with the reconstructed $K(\bn)$ field from CMB-HD may allow for measurement of the He reionization parameters $y_{\rm re}^{\rm He}$ and $\Delta_y^{\rm He}$. 
    The additional redshift information also strengthens H reionization constraints, with all plotted parameters measurable at $\lesssim 1\sigma$ for the same baseline.
    Though CMB-S4 $\times$ LSST shows similar improvements, simultaneously fitting all reionization parameters for both epochs yields $1\sigma$ constraints only for $y_{\rm re}^{\rm H}$.
}
    \label{fig:fisher_results_ellBin}
\end{figure*}

{\bf \em Combined forecasts.} In the second set of forecasts, which account for the proposed cross-correlation with the galaxy-reconstructed radial velocity field, we assume that galaxy density data is available for $\Nzbins = 7$ redshift bins within $0.1 \leq z \leq 5.0$ for both LSST and MegaMapper. 
Maintaining the same assumptions on the CMB experiment specifications used previously for the $KK$-only forecasts, we construct an 11 parameter information matrix that not only includes the ten reionization parameters, but also the bias term $b_{vg}$. 
Figure~\ref{fig:fisher_results_ellBin} summarizes the fractional errors on the fiducial reionization parameters listed in Tab~\ref{tab:set_of_params}. Each of the plotted line styles follows the same conventions as before - solid, dashed, dot-dashed and dotted curves correspond to our forecasts on $y_{\rm re},\, \Delta_y,\, \bar{R},\, \sigma_{\ln R}$, respectively. Once again, the left (right) panel displays the forecasts for the H (He) reionization parameters, with the pink (blue) curves displaying results for th CMB S4 $\times$ LSST (CMB-HD $\times$ MegaMapper) baseline . 

As with the previous set of results, the forecasted parameter errors decrease steadily with increasing $L_{\rm max}$ until $L_{\rm max} \sim 100$, after which the marginal improvement plateaus. 
However, most significantly, the new forecasts indicate that the addition of the $z$-binned $\eta$-field data will allow for measurement of He reionization parameters $y_{\rm re}^{\rm He}$ and $\Delta_y^{\rm He}$ with a baseline matching CMB-HD $\times$ MegaMapper. 
Furthermore, the additional redshift information from the cross-correlation signal improves future prospects of probing H reionization, with all the plotted reionization parameters measurable within $1\sigma$ for the same baseline. 
Although a similar improvement is visible for the forecasts corresponding to CMB-S4 $\times$ LSST, our results indicate that attempting to probe all reionization parameters, across both epochs, while also accounting for possible parameter degeneracies will result in $1\sigma$ constraints only on $y_{\rm re}^{\rm H}$.

A comparison between the forecasted parameter errors in Figs.~\ref{fig:fisher_results_KKonly_ellBin} and \ref{fig:fisher_results_ellBin} confirms that the cross-correlation between the $K(\bn)$ field and the $\eta(\bn, z)$ field is an essential tool in measuring the redshift-evolution of He reionization. 
The improvement in sensitivity to He reionization parameters, driven by the inclusion of the cross-correlation signals $C_L^{K\eta}$ at $z \lesssim 5$, can be attributed to the fact that the additional signals amplify an epoch that specifically corresponds to patchiness in the electron distribution sourced by the gradual (second) ionization of He. 
A similar improvement is reaped by the H reionization parameters because our model of reionization predicts an abundance of non-ionized H roughly matching $\epsilon$ down to $z\sim4$.
The benefits of the cross-correlation are primarily leveraged by the average reionization evolution parameters $y_{\rm re}$ and $\Delta_y$ for both epochs because the additional redshift information available form the $z$-binned cross-correlation with $\eta(\bn, z)$ helps pin down the evolution of $\bxtot(z)$.
Since our bubble distribution $P_X(R)$ (for each epoch) does not evolve with redshift, the cross-correlation helps alleviate degeneracies between $\{\bar{R}^X, \sigma_{\ln R}^X\}$ and $\{y_{\rm re}^X, \Delta_y^X\}$ for both $X = $H and $X = $ He. 
However, even with the addition of the cross-correlation signal, most reionization parameters remain degenerate with bubble-bias terms and therefore $b^{\rm H}$ and $b^{\rm He}$ are once again unmeasurable within $1\sigma$. 

\begin{figure*}
    \centering
    \includegraphics[width=\linewidth]{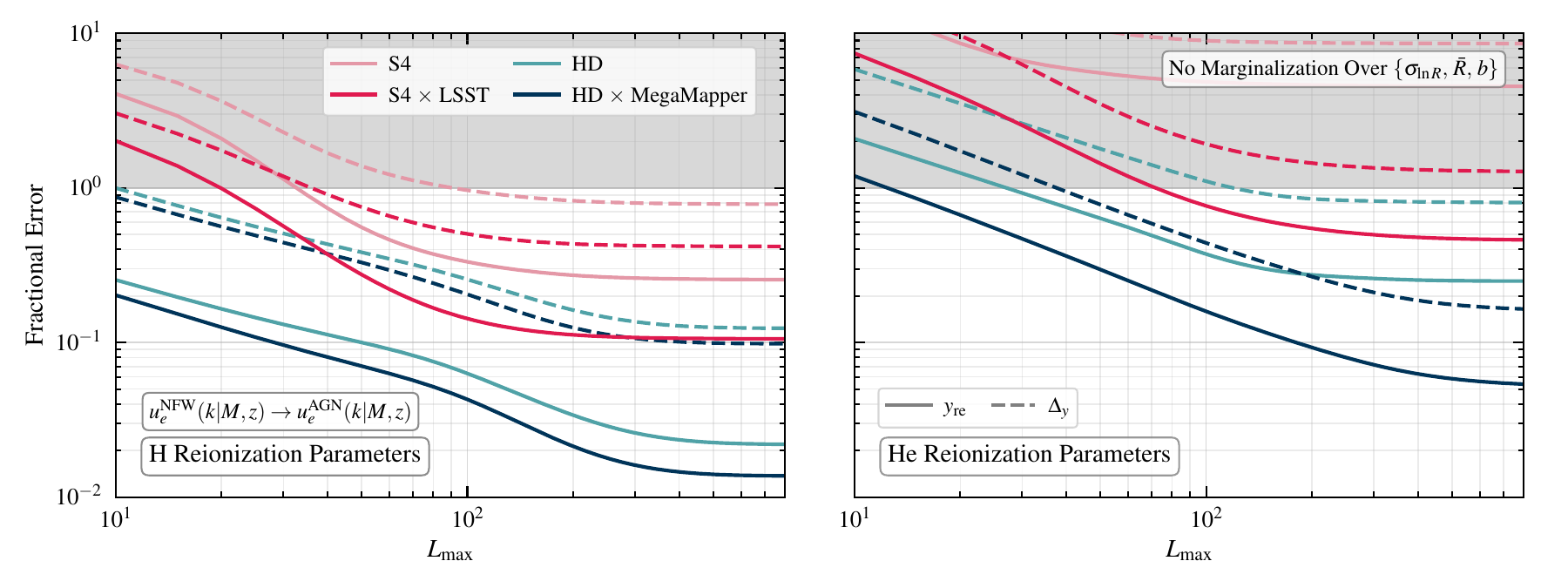}
    \caption{
    Fractional errors on reionization parameters $y_{\rm re}^{X}$ and $\Delta_y^X$ for $X \in \{{\rm H\,(left\, panel)} ,\, {\rm He\,(right\,panel)}\}$ that characterize the redshift evolution of $\bxtot(z)$. The information matrix for these forecasts did not include the patchy-bubble parameters $\bar{R}^{X}$, $\sigma_{\ln R}^X$, and $b^X$ for $X \in \{{\rm H},\, {\rm He}\}$. The light-pink (-blue) curves correspond to forecasts using the $\ell$-binned $KK$ auto-correlation signal from CMB-S4 (CMB-HD) alone. The dark-pink (-blue) curves represents forecasts that \textit{also} account for the [$(\ell\times z)$-binned] $K\eta$ cross-correlation signals from CMB-S4 $\times$ LSST (CMB-HD $\times$ MegaMapper). The cross-correlation forecasts include a marginalization over the bias parameter $b_{vg}$. The redshift evolution of the ionized electron fraction can be determined within $1\sigma$ using the $KK$ signals alone for a CMB-HD-like survey or by leveraging the cross-correlation for either of the baselines considered. Results are computed using the NFW (AGN) profile for $u_e(k|M,z)$ at $z \gtrsim 5$ ($z \lesssim 5$).
    }
    \label{fig:fisher_results_noPatchy_ellBin}
\end{figure*}

{\bf \em Only targeting redshift-evolution.} The discussion above establishes a degeneracy between the patchy bubble parameters $\{\bar{R}^X,\, \sigma_{\ln R}^X,\, b^X\}$ and the redshift evolution parameters $\{y_{\rm re}^X,\, \Delta_y^X\}$ for each $X =$ H and $X = $ He. 
Given this result, it is instructive to consider the ability of upcoming surveys to constrain the set of parameters governing $\bxtot(z)$ \textit{alone}, in the event that simulations of reionization or alternate probes are able to characterize the patchy morphology of each epoch. 
Figure~\ref{fig:fisher_results_noPatchy_ellBin} contains our forecasts on the set of parameters $\{y_{\rm re}^{\rm H},\, \Delta_y^{\rm H},\,  y_{\rm re}^{\rm He},\, \Delta_y^{\rm He}\}$ without marginalization over the bubble parameters. The light-pink (-blue) curves represent forecasts made using the $KK$ auto-correlation signal only, assuming survey specifications that match CMB-S4 (CMB-HD). 
The set of dark-pink (-blue) curves display the same set of forecasts computed assuming access to the $K\eta$ cross-correlation signal, i.e, using the CMB-S4 $\times$ LSST (CMB-HD $\times$ MegaMapper) baseline.  
These forecasts indicate that a baseline similar to CMB-S4 $\times$ LSST will be able to simultaneously constrain the mean redshift of both the epochs and and duration of H reionization. 
Furthermore, with survey specifications corresponding to CMB-HD, the redshift evolution of $\bxtot(z)$ may be measurable within $1\sigma$ with the $KK$ auto-correlation signal alone. 
The inclusion of the cross-correlation with a survey like MegaMapper may allow for $\gtrsim3\sigma$ constraints on all  the displayed parameters.

{\bf\em On the impact of $\ell$-binning.} Having now demonstrated the role of $z$-binned galaxy measurements in alleviating covariances between various reionization parameters, it is important to highlight the significance of $\ell$-binning the kSZ signal in constructing the suite of fields ${K_i(\bn)}$ for $i \in [1,, \Nellbins]$.
As explained in Ref.~\cite{Ferraro:2018izc}, $\ell$-binning serves as a crucial method for reducing degeneracies between parameters characterizing H reionization. While that work specifically discusses its impact on parameters describing $\bxH(z)$, we find that the same effect extends to the parametrization of $P^H(R)$.
Most importantly, leveraging the $\ell$-dependence of the kSZ signal through this method significantly enhances the ability to disentangle the effects of H and He reionization. 
In other words, we find that the covariance between H and He reionization parameters remains minimal---even in forecasts based on the $KK$ auto correlation---largely due to the construction of $\ell$-binned $K$ fields.
Therefore, while the forecasts in this section suggest that neglecting He reionization in parameter estimation may not introduce significant biases in the characterization of $\xH(\br,z)$, this result critically relies on utilizing the $\ell$-dependence of the kSZ effect. 
Without this binning implemented, neglecting He reionization has the potential to bias the constrains on the morphology of H reionization.

{\bf \em On the impact of electron profile.} Given that our model for reionization incorporated into $P_{ee}^{\rm tot}(k,z)$ includes a dependence on $u_e(k|M,z)$, we can assess the impact of small-scale power in $P_{ee}^{\rm tot}(k)$ on the parameter-error forecasts presented in this section. 
To explore this effect, we vary the late-time ($z\lesssim 5$) electron profile $u_e^Y(k|M, z)$ across $Y \in \{W_e(k)\times{\rm NFW},\, {\rm AGN},\, {\rm NFW}\}$, while holding the early-time profile to an NFW.
Consistent with our previous discussion regarding the impact of $u_e(k|M, z)$ on forecasted SNRs, we find that the changing between the NFW and AGN profiles at late times does not significantly impact fractional-error forecasts on any of the reionization parameters. 
Furthermore, the choice of $W_e(k)\times{\rm NFW}$ at late times causes the most prominent change in the fractional error forecasts. 
Despite the decrease in the $C_L^{KK}$ and $C_L^{K\eta}$ due to the overall decrease in small-scale power sourcing the kSZ effect, we find that choosing the $W_e(k)\times{\rm NFW}$ at low redshifts leads to smaller forecasted fractional errors on H reionization parameters. 
The improvement across the parameter space is bounded by a factor of $\sim 2$ for both $\VEV{KK}$-only and $\VEV{KK} + \VEV{K\eta}$ forecasts. 
The enhanced sensitivity to H reionization is explained by the decrease in noise $N_L^{KK}$ as well as a diminished contribution from $P_{ee}^{\rm tot}$ to the auto- and cross-correlation signals.
In contrast, we find that the choice of $W_e(k)\times{\rm NFW}$ at late times decreases experiment sensitivity to He reionization. 
This diminished sensitivity is a direct consequence of the decreased power in terms $\propto \VEV{\xHe\netot}_k$ that contribute to $P_{ee}^{\rm ion}(k,z)$ and carry information about the patchy morphology of He reionization. 
The degradation in forecasted errors is, however, smaller than the improvement seen with H reionization parameters---the change is never larger than a factor of $\sim 1.5$ for both $\VEV{KK}$-only and $\VEV{KK} + \VEV{K\eta}$ forecasts. 
Finally, we find that changes to the assumed early-time electron distribution do not visibly affect any forecasted parameter errors. 

Finally, it is instructive to note that the parametrization of models characterizing both epochs of reionization exactly matches the model used in Ref.~\cite{Caliskan:2023yov}. 
Furthermore, the experiment specifications assumed for each baseline considered in the forecasts presented here are similar to those assumed in Ref.~\cite{Caliskan:2023yov}, except for the replacement of LSST with MegaMapper for the CMB-HD-based cross-correlation forecasts. 
As a result, the fractional parameter errors in this section, particularly those displaying the power of the cross-correlation with galaxy survey measurements, can be directly compared with the results presented in Sec. IV of Ref.~\cite{Caliskan:2023yov}.
Although the kSZ-based auto- and cross-correlation trispectrum signals present a slightly weaker sensitivity to bubble parameters $\{\bar{R},\,\sigma_{\ln R},\, b\}$ for both epochs, the set of $\{K_i(\bn)\}$ and $\{\eta_\alpha\}$ fields do not need to be reconstructed to small angular scales to obtain competitive constraints on the midpoint and duration of both H and He reionization. 

\section{Discussion}
\label{sec: Conclusions}
In this paper, we consider the sensitivity of the kSZ-based trispectrum statistic to the epochs of H and He reionization.  
While the epoch of H reionization ($z \sim 8.5$), driven by early star formation, has been studied extensively, its He counterpart ($z \sim 3$) has garnered less attention despite its close link to the properties of early quasar populations and the formation of AGN~\citep[e.g,][]{LaPlante2017,LaPlante:2015rea,LaPlante2018}.
Theoretical models and forecasts on He reionization~\citep[e.g.,][]{Furlanetto:2007gn, McQuinn2009, Kapahtia:2024rgw, Hotinli:2022jna, Hotinli:2022jnt, Caliskan:2023yov, LaPlante:2015rea, LaPlante:2016bzu, LaPlante:2017xzz} have indicated that a joint analysis of both epochs holds the promise of a more coherent understanding of the formation and evolution of early ionizing sources. 
Given this motivation, in this paper we first extend the application of the kSZ trispectrum statistic---introduced in Ref.~\cite{Smith:2016lnt} as a probe of H reionization---to incorporate the effects of patchy He reionization and assess the probe's combined sensitivity to both epochs.

Despite the established sensitivity of the kSZ trispectrum to the epoch of H reionization, the first joint forecasts on both epochs---using optical-depth fluctuations imprinted in the CMB signal (Ref.~\cite{Caliskan:2023yov})---indicate that accessing the morphology of He reionization using a $z$-integrated probe will likely be unfeasible, owing to He's low relative abundance. Moreover, Ref.~\cite{Caliskan:2023yov} tackles this challenge by incorporating low-redshift cross-correlations to enhance experiment sensitivity to this latter epoch.
Anticipating a similar suppression in the $C_L^{KK}$ signals, we propose a novel cross-correlation of the kSZ-weighted temperature-squared maps with low-redshift galaxy survey measurements, aimed at amplifying the signal from He reionization relative to its H counterpart.

In this work, we begin with a derivation of  $P_{ee}^{\rm ion}(k,z)$ through both the epochs of reionization, rooted in the HOD prescription described in Ref.~\cite{Mortonson:2006re}. 
The generalized formalism we present allows for independent modeling of the two (possibly overlapping) epochs of reionization, leaving room for more quasar-measurement-driven models of He reionization.  
We also present a derivation of the more condensed model ultimately adopted in the final forecasts, clarifying all simplifying assumptions made.  
Although this reionization model closely follows the form used in previous forecasts based on other CMB-secondary probes ~\citep[e.g.,][]{Dvorkin:2008tf, Caliskan:2023yov}, we extend our final expression for $P_{ee}^{\rm ion}(k,z)$ to include an explicit dependence on the small-scale electron profile within halos.  
This extension allows our forecasts to inform how different HOD models could affect future measurement SNRs and parameter errors.

Given this model for $P_{ee}^{\rm ion}(k,z)$, we proceed to compute the suite of $\ell$-binned trispectrum signals $C_L^{KK}$ using the prescription presented in Refs.~\cite{Smith:2016lnt, Ferraro:2018izc}.
We extend prior analysis and forecasts that have leveraged the $KK$ trispectrum signal, by not only including the effects of He reionization, but also incorporating a dependence on `bubble' parameters, that characterize the size-distribution and clustering of ionized regions in each epoch (similar to the analyses presented in Refs.~\cite{Dvorkin:2008tf, Caliskan:2023yov}).
We then derive expressions for the proposed cross-correlation between the $\ell$-binned kSZ-squared fields $\{K(\bn)\}$ used in the trispectrum statistic and the $z$-binned radial-velocity fields (squared) $\{\eta(\bn)\}$ reconstructed from galaxy survey measurements. 

With the signal and noise defined, we present forecasts for our ability to measure the auto- and cross-correlation signals, as well as to jointly characterize the patchy morphology of each epoch.
We consider two separate baselines for our forecasts---the first representing the cross-correlation of kSZ measurements from a CMB-S4-like survey~\citep{Abazajian:2019eic} with galaxy-survey measurements from LSST~\citep{2009arXiv0912.0201L}, and the second corresponding to data from a CMB-HD-like experiment~\citep{CMB-HD:2022bsz} cross-correlated with the upcoming MegaMapper survey (a planned  proposed spectroscopic follow-up to LSST)~\citep{Schlegel:2022vrv}. 
It is important to note that, since the original ``MegaMapper'' concept has evolved into the broader Stage-5 spectroscopic survey (Spec-S5) framework, we take MegaMapper to represent a plausible Spec-S5-like survey~\cite{Spec-S5:2025uom}. 

The forecasted measurement SNRs show that the CMB-HD $\times$ MegaMapper baseline can detect clustering in the $K(\bn)$ field sourced by large-scale correlations in the locally measured kSZ power with an SNR of $\sim 100$ at $L_{\rm max} \sim 50$ for the forecasts that include both $C_L^{KK}$ and the proposed $C_L^{K\eta}$. 
This indicates that the fields do not have to be reconstructed to small scales for a detection of the signal. 
We also forecast the specific contributions of H and He reionization to the total detection SNR, for both the auto- and cross-correlation signals. 
These results corroborate the initial hypothesis that the contribution from He reionization, as a fraction of the total SNR, will be distinctly suppressed in the auto-correlation $C_L^{KK}$ measurement. 
Moreover, as a fraction of the total measurement SNR, the contribution from He reionization to the cross-correlation SNR is significantly increased, confirming that the cross-correlation plays a vital role in amplifying the signal from the latter epoch. 

Forecasts on fractional parameter errors (as a function of $L_{\rm max}$) from the joint information-matrix analysis, accounting for reionization parameters across both epochs, show that one can constrain both the redshift evolution and one of the parameters characterizing the size distribution of ionized regions during H reionization using $C_L^{KK}$ measurements alone, from a CMB-HD-like survey.  
However, as anticipated from the forecasted SNRs, all He reionization parameters remain unconstrained.
When the cross-correlation signals $C_L^{K\eta}$ are included, the additional redshift-binned information at $z \lesssim 5$ amplifies the signal from He reionization, enabling $1\sigma$ constraints on its redshift evolution using the CMB-HD $\times$ MegaMapper baseline.  
Moreover, the inclusion of cross-correlation information improves the H reionization constraints as well, with all associated parameters measurable within $1\sigma$ by the same baseline.  
Finally, we find that if the experiments target only the mean redshift and duration of each epoch---without marginalization over the patchy morphology of each epoch---the CMB-HD $\times$ MegaMapper baseline can place $\lesssim 3\sigma$ constraints on all four parameters for $L_{\rm max} \sim 200$.

It is important to acknowledge that the results presented in this paper rely on a simplified model of reionization for both H and He.
This choice was made not only to facilitate comparison with previous forecasts on these epochs~\citep[e.g.,][]{Dvorkin:2008tf, Caliskan:2023yov} but also to provide forecasts that illustrate how the size of ionized regions or redshift evolution might impact future constraints.
Furthermore, given the weaker sensitivity of the (small-scale averaged) $K(\bn)$ probe to bubble parameters, a more complex model would likely not be significantly more instructive for initial forecasts.
However, the same two-data sets can be cross-correlated to construct a bispectrum-level statistic---using the temperature-squared $\{K(\bn)\}$ fields and galaxy density measurements $\delta_g(\bk,z)$---that will likely be more sensitive to the small-scale patchiness of the later epoch. 
This type of cross-correlation may more effectively exploit the fact that galaxies serve not only as large-scale tracers of the peculiar velocity field experienced by ionized electron distributions but are also intrinsically linked to the distribution of ionizing sources during He reionization. As a result, this correlation yields a nonzero $\VEV{\xHe\delta_g}_k$ signal across both $1h$ and $2h$ scales.
This cross-correlation might, therefore, have a better sensitivity to the small-scale physics governing He reionization, possibly instructing on clustering properties and lifetimes of quasars. 
Therefore, we leave the development of an astrophysically motivated model for He reionization, along with forecasts utilizing this alternative cross-correlation probe, for our upcoming work.

Finally, our forecasts do not include the effect of some sources of reconstruction noise in both $\{K_i(\bn)\}$ and $\{\eta_\alpha(\bn)\}$.
Observations of the large-scale clustering of locally measured CMB power will include, for example, a $C_L^{KK}$ signal sourced by power along the CMB line-of-sight whose amplitude is linear in the matter density field. Reference~\cite{Smith:2016lnt} establishes that this effect is likely sub-dominant relative to the kSZ sourced $C_L^{KK}$. The tSZ effect is known to contribute a constant offset to the observed $KK$ signal at large $L$. Furthermore, late-time lensing of the CMB is known to induce a $C_L^{KK}$ signal with power comparable to the reconstruction noise $N_L^{KK}$. References~\cite{MacCrann:2024ahs, SPT-3G:2024lko} detail the magnitude of these effects and assesses various techniques for their removal. 
Given the large dimensionality of our parameter space, including marginalization over multiple bias parameters, we do not account for biases in measurements of the kSZ sourced $C_L^{KK}$ caused by the aforementioned foregrounds. 
Furthermore, on the front of velocity reconstruction from galaxy-survey measurements, it is important to note that LSST provides galaxy redshifts based on photometry, which are subject to photo-$z$ uncertainties, whereas MegaMapper is designed to obtain high-precision spectroscopic redshifts for a subset of LSST galaxies, especially at higher redshifts.
However, we do not include the effects of photo-$z$ errors in our forecasts. 
Therefore, our comparison between LSST and MegaMapper focuses on differences in galaxy number density rather than redshift precision. 
As such, the relative gain in SNR from using MegaMapper over LSST is likely underestimated here; in principle, LSST’s photo-$z$ errors would degrade the reconstruction fidelity, and a spectroscopic survey like MegaMapper should offer a more significant improvement.
A comprehensive analysis of these foreground effects and other possible sources of reconstruction noise, particularly for He reionization measurements, is left for future work.

In conclusion, our forecasts demonstrate that the upcoming generation of high-resolution CMB experiments coupled with the forthcoming galaxy surveys have the potential to transform the kSZ effect from a qualitative indicator of H reionization into a tool that can constrain the time, duration, and morphology of ionizing sources through the Universe's evolution at $z\lesssim 15$.
While the auto-correlation of kSZ-weighted temperature-squared maps alone delivers competitive measurements of H reionization parameters, with cross-correlations that leverage tomographic measurements of the low-redshift cosmological velocity fields, future surveys may obtain constraints on the redshift evolution of He reionization as well.
This combined analysis has the potential to offer a more complete picture of the evolution and morphology of early ionizing photons. 
As the velocity reconstruction methods mature and complementary probes such as 21-cm tomography and line-intensity mapping become available, the observational landscape will support far richer models that trace the sources of reionization and thermal feedback self-consistently.
The framework introduced here also emphasizes a vital point: cross-correlating three-dimensional tracers of large-scale structure, with CMB secondaries is not only advantageous but also crucial for extracting the full cosmological  narrative encoded in the small-scale CMB sky.

\acknowledgments
We are grateful to Paul La Plante, Matthew Johnson, Amalia Madden, and Sam Joseph Goldstein for useful discussions.
This work was supported at Johns Hopkins University by NSF Grant No.\ 2412361, NASA ATP Grant No.\ 80NSSC24K1226, the Guggenheim Foundation, and the John Templeton Foundation.  MK thanks the Center for Computational Astrophysics at the Flatiron Institute and the Institute for Advanced Study for hospitality.
S.C.H.~was supported by the P.~J.~E.~Peebles Fellowship at Perimeter Institute for Theoretical Physics and the Horizon Fellowship from Johns Hopkins University. This research was supported in part by Perimeter Institute for Theoretical Physics. Research at Perimeter Institute is supported by the Government of Canada through the Department of Innovation, Science and Economic Development Canada and by the Province of Ontario through the Ministry of Research, Innovation and Science. 
This work was in part carried out at the Advanced Research Computing at Hopkins (ARCH) core facility (rockfish.jhu.edu), which is supported by the NSF Grant No.~OAC1920103.

\appendix

\section{Total Electron Distribution Profile Specifications}
\label{appendix: Total Electron Distribution Profile Specifications}
For the forecasts in this paper, we use the halo model to
calculate the non-linear ionized electron power-spectrum $P_{ee}^{\rm ion}(k,z)$. 
This calculation involves the computation of the set of two point functions $\{\VEV{x_Xx_{X'}}_k,\, \VEV{x_X\netot}_k,\, \VEV{\netot\netot}_k \}$ for $X,\,X' \in \{$H, He$\}$. 
As suggested by Eqs.~\eqref{eq: ion_elec_PS_as_ft_2pt_terms}-\eqref{eq: ne_tot_x_X_fourier_two_point}, this calculation requires specification of the halo mass function $n(M)$, the halo bias $b(M)$, as well as the assumed distribution of \textit{all} electrons on small-scales within halos $u_e(k|M,z)$. 
These ingredients are detailed below. 

The matter over-density field $\delta_m(\bk,z)$ smoothed by the top-hat window function, is calculated as follows:
\begin{eqnarray}
    \delta_m^W(\br,z) = \int \dd^3\br' \delta_m(\br', z)W_R(\br - \br')\,,
\end{eqnarray}
where the well-known, top-hat window function takes the following form in Fourier-space:
\begin{eqnarray}
    W_R(k) = \frac{3}{\lr{(}{kR}{)}^3} \lr{[}{\sin(kR) - kR\cos(kR)}{]}. 
\end{eqnarray}
Given this definition, the rms variance of mass within a sphere of radius $R$, that contains mass $M = 4\pi R^3\rho_m/3$, is defined as:
\begin{equation}
    \sigma^2(M, z) = \frac{1}{2\pi^2}\int_{0}^{\infty} \dd k\ \ k^2\  P_{mm}^{\rm lin}(k,z)W_R(k)^2.
\end{equation}
This quantity can then be used to define the halo-mass function:
\begin{equation}
    n(M, z) = f(\sigma,z) \frac{\rho_{m}}{M^2}\frac{d\ln[\sigma(M,z)^{-1}]}{d\ln(M)}\,,
\end{equation}
where $f(\sigma,z)$ denotes the collapse fraction. For the signals and associated forecasts presented in this paper, we assume use the Tinker collapse fraction~\cite{Tinker:2008ff},
\begin{equation}
    f(\sigma, z) = A\Bigg[\Big(\frac{\sigma}{b}\Big)^{-a} + 1\Bigg]e^{-c/\sigma^2},
\end{equation}
with $A=0.186$, $a = 1.47$, $b = 2.57$, and $c=1.19$. The linear halo-bias consistent with the above collapse fraction is:
\begin{equation}
    \begin{split}
        b_h(M,z) = 1 + \frac{1}{\sqrt{a}\delta_c} \Bigg[ &\sqrt{a}(a\nu^2) + \sqrt{a}b(a\nu^2)^{1-c} \\
        &- \frac{(a\nu^2)^c}{(a\nu^2)^c + b(1-c)(1-c/2)} \Bigg]\,,
    \end{split}
\end{equation}
where $\nu(m,z) \equiv \delta_c/ \sigma(M,z)$, and the fit parameters are $a=0.707$, $b = 0.5$, and $c = 0.6$~\cite{Tinker:2010my}. Note that the above set of expressions approximately satisfy the following consistency relation:
\begin{equation}
    \int_{-\infty}^{\infty}\dd\ln{M}\ M^2n(M,z)\Big(\frac{M}{\rho_m}\Big)b_h(M,z) = 1.
\end{equation}

With the distribution of halos determined via the above formalism, the final step in characterizing the clustering of electrons on small scales is defining the assumed distribution profile of electrons within spherical halos. 

\begin{figure}
    \centering
    \includegraphics[width=\linewidth]{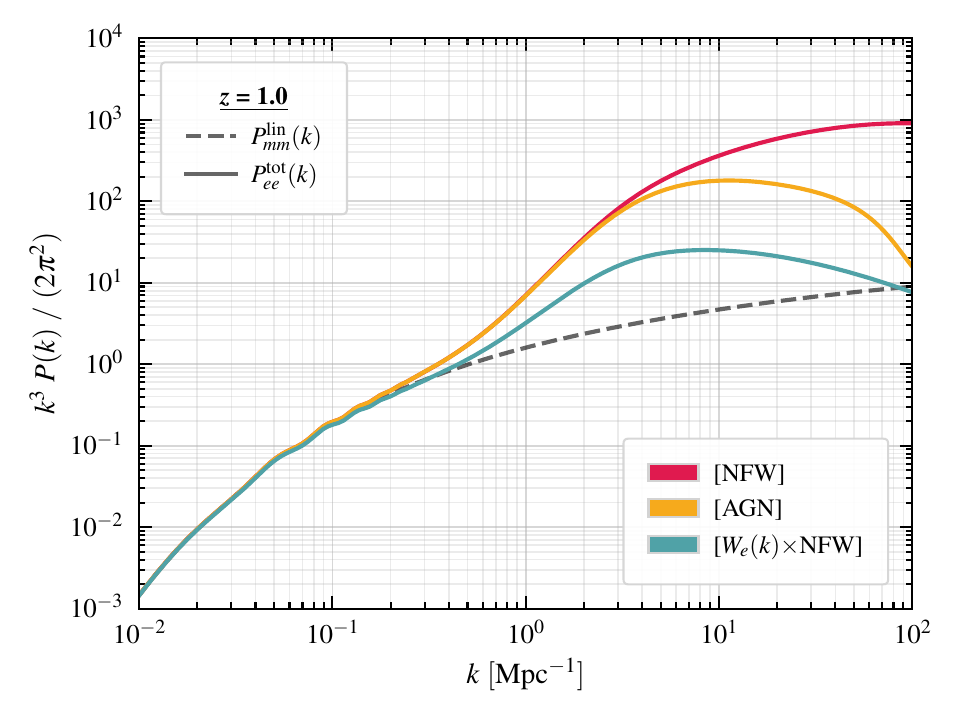}
    \caption{Total electron power spectrum $P_{ee}^{\rm tot}(k,z)$ [$\VEV{\netot\netot}_k/(\bnetot)^2$] at $z = 1.0$. 
    Each colored line corresponds to a specific small-scale electron-distribution profile considered in the final forecasts. 
    Since the profile-dependency only appears in the 1$h$ term of $\VEV{\netot\netot}_k$ [Eq.~\eqref{eq: tot_elec_dist_fourier_2pt}], the power-spectra only deviate from each other at small $k$. 
    The linear-matter power spectrum $\Plin(k)$ is also plotted in the gray dashed curve for comparison.
    }
    \label{fig:total_elec_PS_no_re}
\end{figure}

In our forecasts, we consider three separate models for the Fourier-space distribution profile of electrons that predict differing power in the total electron power spectrum $P_{ee}^{\rm tot}(k,z)$ on small scales. In the first scenario, we assume that the electrons are distributed according to an NFW profile which, in real space, takes the following form:
\begin{equation}
    \rho_e^{\rm NFW}(r|M,\, z) = \frac{\rho_s}{(r/R_s) (1 + r/R_s)^2}\,,
    \label{eq: NFW_profile}
\end{equation}
where $R_s$ is the scale radius, related to the virial radius $R_{200}(M)$ of a halo via the concentration parameter $c \equiv R_{200}/R_s$. The halo mass and redshift dependence of the above profile are therefore induced by the model for the concentration parameter. For our forecasts, we assume
\begin{equation}
    c(M,z) = A \Big(\frac{M}{2\times 10^{12}\ h^{-1} M_{\odot}}\Big)^{\alpha}(1+z)^\beta,
\end{equation}
where $h$ is the reduced Hubble constant and we use the fit parameters $A = 7.85$, $\alpha = -0.081$, and $\beta = -0.71$~\cite{Smith:2018bpn}.

The second model considered in our forecasts, called the AGN model, corresponds a fitting function from Ref.~\cite{Battaglia:2016xbi} based on simulations with the `AGN' sub-grid feedback model given by:
\begin{equation}
    \rho_{\text{gas}} = \frac{\Omega_b}{\Omega_m}\rho_{\text{c}}(z)\bar{\rho}_0\Big(\frac{x}{x_c}\Big)^\gamma\Bigg[1 + \Big(\frac{x}{x_c}\Big)^\alpha\Bigg]^{-\frac{\beta-\gamma}{\alpha}},
\end{equation}
where we have dropped the explicit dependence of some of the above parameters on mass and redshift for ease of notation. In the above model, $x = r/R_{200}(m, z)$, $\gamma = -0.2$ and $x_c = 0.5$. The remaining parameters $\bar{\rho}_0(m,z),\, \alpha(m,z)$, and $\beta(m,z)$ are fitted with a power law in halo mass and redshift:
\begin{equation}
    A = A_0^x\Bigg(\frac{m}{10^{14} M_\odot}\Bigg)^{\alpha_m^x}(1+z)^{\alpha_z^x},
\end{equation}
where the parameters for the AGN model used in this paper have been lifted from Table 2 of Ref. \cite{Battaglia:2016xbi}.

Finally, the third model considered in our discussions is lifted from the approximate late-time electron power spectrum used in Ref.~\cite{Smith:2016lnt}. Reference~\cite{Smith:2016lnt} approximates the late-time electron power spectrum as $W_e(k,z)^2P_{mm}^{\rm nl}(k,z)$, where $P_{mm}^{\rm nl}$ is the late-time non-linear matter power spectrum and $W_e(k,z)$ is defined as:
\begin{equation}
    W_e(k,z) \equiv \lr{(}{1 + \frac{kD(z)}{0.5 h\,{\rm Mpc^{-1}}}}{)}^{-1/2},
\end{equation}
where $D(z)$ is the growth function normalized to $1/(1+z)$ at high $z$. 
Reference~\cite{Smith:2016lnt} explains that this form of $W(k,z)$ is a simple fitting function which is approximately consistent with the `cooling + star formation model' from Ref.~\cite{Shaw:2011sy} for both $C_\ell^{\rm kSZ}$ and $\dd C_\ell^{\rm kSZ}/ \dd z$ from late times. 

In our forecasts, we assume that the non-linear matter power-spectrum is characterized on small-scales by the NFW profile described above. 
In other words, our model for the $u_e^{W_e(k)\times{\rm NFW}}$ is calculated as $W_e(k)\times u_e^{\rm NFW}$, where $u_e^{\rm NFW}$ is the Fourier-space counterpart of the profile presented in Eq.~\ref{eq: NFW_profile} (normalized to one at low $k$). 
Note that we neglect the prefactor of 0.85 used in Ref.~\cite{Smith:2016lnt} because our model for ionization assumes that, after the completion of H and He reionization, \textit{all} electrons are ionized. 

Figure~\ref{fig:total_elec_PS_no_re} displays the total electron power-spectrum ($\propto\VEV{n_e^{\rm tot}n_e^{\rm tot}}_k$) for each of the above discussed models for the electron profile $u_e(\bk|M,z)$. All three models require that the electron gas traces dark-matter on large scales, and as a a result, each of the computed power-spectra deviate from $P_{mm}^{\rm lin}$ only on small scales. The NFW profile results in the largest power on small scales, with the AGN profile predicting a slightly lower power and the $W_e(k)\times$NFW profile predicting the smallest power at large $k$. 

\section{Calculation of $P_{\eta\eta}^{\perp}$ Integral}
\label{appendix: Calculation of P_eta Integral}
In this section, we briefly summarize the methodology adopted in this work to compute the power-spectrum of the $\eta(\bn,z)$ field for modes $\bk$ that are perpendicular to the line of sight ($k_r = 0$). We will focus on the calculation of the integral in Eq.~\eqref{eq: p_eta_perp}, however the methodology is easily extended to the calculation of the $\eta$-field reconstruction noise $N_{\eta\eta}(k,z)$ in Eq.~\eqref{eq: N_eta_perp}.

The integrand in Eq.~\eqref{eq: p_eta_perp} calls for the decomposition of two vectors --- $\bk = (0,\, \bk_\perp)$ and $\bk' = (\mu k',\, \sqrt{1-\mu^2}k'\hat{\bk}'_\perp)$ --- where $\mu$ is the cosine of the angle between $\bk'$ and the line of sight $\bn$, $\hat{\bk}'_\perp$ is the unit-normalized component of $\bk'$ perpendicular to the line of sight, and we have imposed the reconstruction of modes with $k_r = 0$. Under this decomposition, we have:
\begin{equation}
    |\bk-\bk'| = \lr{[}{k^2 + (k')^2 - 2kk'\sqrt{1-\mu^2}\cos\phi}{]}^{1/2}\,,
\end{equation}
where $\phi$ is assumed to be the angle between $\bk_\perp$ and $\bk'_\perp$. The integral can then be simplified to
\begin{align}
    P_{\eta\eta}^\perp(k) = \frac{2}{\VEV{v_r^2}^2}
    \int_{-1}^{1}&\frac{\dd y}{\sqrt{1-y^2}}
    \int_{-1}^1\dd\mu
    \int \frac{\dd k'}{(2\pi)^3} \notag \\
    &\quad\times \frac{\mu^4(k')^4}{|\bk - \bk'|^2}
    P_{vv}(k')P_{vv}(|\bk-\bk'|)\,,
\end{align}
where $y\equiv \cos{\phi}$ and the redshift dependence of $\VEV{v_r(z)^2}^2$ and all the power-spectra has been suppressed for ease of notation.

\bibliography{draft}

\end{document}